\documentclass[12pt,a4paper]{article}
\usepackage{amsmath,caption,bbm}
\usepackage[top=1.2in, bottom=1.2in, left=1.1in, right=1.1in]{geometry}
\usepackage[inline]{enumitem}
\DeclareCaptionStyle{italic}[justification=centering]{labelfont={bf},textfont={it},labelsep=colon}
\usepackage{graphicx,psfrag,epsf}
\usepackage{enumerate}
\usepackage{multirow}
\usepackage{natbib}
\usepackage{url} 
\usepackage{booktabs, subfig ,bm, bbm, paralist,mathpazo,tikz,longtable,microtype,dsfont,rotating} 
\usepackage[pdftex,colorlinks=true]{hyperref}
\definecolor{darkblue}{rgb}{0,0,.6}
\hypersetup{citecolor=darkblue,linkcolor=darkblue,urlcolor=darkblue}
\newcommand{\X}{\mathcal{X}}
\newcommand{\cov}{\text{cov}}

\newcommand{\blind}{0}

\newcommand{\norm}[1]{\left\lVert#1\right\rVert}
\newcommand{\E}{\text{E}}

\graphicspath{{plots/}}
\DeclareMathOperator*{\argmin}{\arg\!\min}
\newsavebox\CBox
\def\textBF#1{\sbox\CBox{#1}\resizebox{\wd\CBox}{\ht\CBox}{\textbf{#1}}}

\definecolor{a0}{rgb}{0.0, 0.5, 0.0}
\definecolor{bistre}{rgb}{0.24, 0.17, 0.12}
\definecolor{amethyst}{rgb}{0.6, 0.4, 0.8}
\definecolor{blue-violet}{rgb}{0.54, 0.17, 0.89}
\definecolor{Rcolor}{RGB}{150,160,190}
\definecolor{blush}{rgb}{0.87, 0.36, 0.51}
\definecolor{brightturquoise}{rgb}{0.03, 0.91, 0.87}
\definecolor{burntorange}{rgb}{0.8, 0.33, 0.0}

\date{\today}
\AtBeginDocument{}

\begin{document}

\def\spacingset#1{\renewcommand{\baselinestretch}
{#1}\small\normalsize} \spacingset{1}

\if0\blind
{
  \title{\bf Intraday forecasts of a volatility index: Functional time series methods with dynamic updating}
  \author{Han Lin Shang\thanks{Postal address: Research School of Finance, Actuarial Studies and Statistics, Level 4, Building 26C, Australian National University, Kingsley Street, Canberra, ACT 2601, Australia; Telephone: +61(2) 612 50535; Fax: +61(2) 612 50087; Email: hanlin.shang@anu.edu.au.},\quad Yang Yang,
  \hspace{.2cm}\\
    Research School of Finance, Actuarial Studies and Statistics \\
    Australian National University \\
    \\
    Fearghal Kearney \\
    Queen's Management School \\
    Queen's University Belfast 
}
  \maketitle
} \fi

\if1\blind
{
  \bigskip
  \bigskip
  \bigskip
  \begin{center}
    {\LARGE\bf Title}
\end{center}
  \medskip
} \fi

\bigskip

\def\spacingset#1{\renewcommand{\baselinestretch}
{#1}\small\normalsize} \spacingset{1}
\spacingset{1.45}

\begin{abstract}
As a forward-looking measure of future equity market volatility, the VIX index has gained immense popularity in recent years to become a key measure of risk for market analysts and academics. We consider discrete reported intraday VIX tick values as realisations of a collection of curves observed sequentially on equally spaced and dense grids over time and utilise functional data analysis techniques to produce one-day-ahead forecasts of these curves. The proposed method facilitates the investigation of dynamic changes in the index over very short time intervals as showcased using the 15-second high-frequency VIX index values. With the help of dynamic updating techniques, our point and interval forecasts are shown to enjoy improved accuracy over conventional time series models.
\vspace{20pt}

\end{abstract}


\newpage
\section{Introduction}\label{sec:intro}

In 1993, the Chicago Board Options Exchange (CBOE) introduced the CBOE  volatility index (VIX) implied by S\&P 500 Index option prices. VIX is computed based on a weighted portfolio of 30-day S\&P 500 calls and puts, to construct a forward-looking measure of future equity market volatility. Driven by the introduction of a range of VIX derivatives: futures, options, and exchange-traded notes, the VIX index has gained immense popularity in recent years. It has become a key measure of risk for market analysts and academics alike, causing it to be referred to as the `investor fear gauge' by the financial press and investment community \citep[see, e.g.,][]{Whaley00}. Therefore, producing accurate forecasts of future VIX index values are of great importance to risk management. Such forecasts are of interest to all parties involved in analysing VIX and VIX-related instruments, including hedge and pension funds, endowments, and even retail investors.

While producing volatility forecasts of the S\&P 500 and other equity indices is popular in the literature \citep[see, e.g.,][]{WZ10, Xu99, BT07, KU16}, modelling implied volatility indices have received less research attention. Researchers have previously considered VIX index values as discrete time series, producing forecasts using autoregressive fractionally integrated moving average models and heterogeneous autoregressive models \citep[see, e.g.,][]{KST08, FMS14}. \cite{PS16} extend these approaches by successfully adopting them in a hybrid genetic algorithm support vector regression of implied volatility index data \citep[see, e.g.,][for further details]{DLKST13, SSTK14}. Besides the drawback of the studies being conducted at a daily frequency, another common disadvantage is that the discrete models employed ignore the underlying implied VIX evolution dynamics, that is, how the index moves from time $t-1$ to time $t$. Therefore, the underlying stochastic process that generates intraday VIX observations cannot be determined. However, functional data analysis techniques, such as the techniques we implement in the present study, can be used to extract additional information beneath a time series of functions, assisting the investigation of essential sources of pattern and variation \citep[see, e.g.,][Chapter 1]{HK12, KR17, RS06}.

Functional time series is attracting an ever-increasing focus, leading to a rapidly growing body of research on modelling and forecasting. From a parametric perspective, \cite{Bosq00} proposed the functional autoregressive of order 1 (FAR(1)) and derived one-step-ahead forecasts that are based on a regularised form of the Yule-Walker equations. FAR(1) was later extended to a FAR($p$), under which order $p$ can be determined via \cite{KR13}'s hypothesis testing procedure. \cite{ANH15} introduced a vector autoregressive (VAR) model based on the functional principal component analysis (PCA), together with a forecasting method based on VAR forecasts of principal component scores. This VAR model can also be considered as an extension of \cite{HS09}, where principal component scores are forecast by a univariate time series forecasting method, such as the autoregressive integrated moving average (ARIMA) method. \cite{KK16} proposed the functional moving average (FMA) process with an innovations algorithm to obtain the best linear predictor. \cite{KKW17} extended the FAR and FMA processes to a vector autoregressive moving average model, which can be considered a more straightforward estimation approach of the functional autoregressive moving average (FARMA) model. Recently, \cite{LRS17} considered long-range dependent curve time series and proposed a functional autoregressive fractionally integrated moving average model. From a nonparametric perspective, \cite{BCS00} introduced functional kernel regression to model the temporal dependence by a similarity measure characterised by semi-metric, bandwidth, and kernel functions. From a semi-parametric perspective, \cite{AV08} proposed a semi-functional partial linear model that combines parametric and nonparametric models, enabling consideration of additive covariates and the use of a continuous path in the past to predict future values of a stochastic process. Apart from the estimation of a conditional mean, \cite{HHR13} consider a functional analogy of the autoregressive conditional heteroscedasticity model for modelling conditional variance. Recently, \cite{KRS17} proposed a portmanteau test for measuring autocorrelation under a functional analogue of the generalised autoregressive conditional heteroscedasticity model.

Among the various modelling and forecasting methods for functional time series, such as those listed above, many adopt functional PCA as a dimension reduction tool. Functional PCA can summarise an infinite-dimensional object into finite dimensions, sacrificing little information in the process. We utilise the functional PCA approach, also considered in \cite{HS09} and \cite{ANH15}, to decompose a time series of functions into a set of functional principal components and their corresponding principal component scores. Correlation among each set of principal component scores obtained in functional PCA decomposition possesses temporal dependence information of the original functional time series. To forecast principal component scores, \cite{HS09} considered a univariate time series forecasting method (e.g., autoregressive integrated moving average), while \cite{ANH15} apply a multivariate time series forecasting method (e.g., a vector autoregressive model). Conditioning on the past curves and estimated functional principal components, point forecasts can be obtained by multiplying the forecast principal component scores by the estimated functional principal components. This type of algorithm is referred to as the `TS method' in the remaining sections because it relies on either univariate or multivariate time series forecasting methods.

This paper is closely related to a previous study involving forecasting five-minute S\&P 500 index returns using various functional time series methods with dynamic updating by \cite{Shang17B}. \cite{Shang17B} noted that applying the forecasting methods to intraday financial data with denser time intervals (less than 5 minutes) would be a validation of the proposed forecasting methods. We provide this validation through the use of the higher resolution 15-second VIX data. Our study contributes further by utilising robust approaches that control for the destabilising effect of outliers, and by extending the application of functional techniques to further the understanding of intraday implied VIX movements. We uncover intraday VIX patterns that allow academics and practitioners to better prepare for extreme events and given the proliferation of VIX-related futures, options, and exchange-traded products. While it is not our primary focus, our research could form the basis of idiosyncratic high frequency hedging and trading strategies.

The rest of this paper is structured as follows. In Section~\ref{sec: VIX}, we describe the VIX data employed in our study. In Section~\ref{sec: methodology}, we introduce the functional PCA, the robust functional PCA, and the robust regularised singular value decomposition. We present several dynamic updating methods for updating point forecasts in Section~\ref{sec: updating} and introduce some dynamic updating methods for updating interval forecasts in Section~\ref{sec: interval}. Using the forecast error measures shown in Section~\ref{sec: accuracy}, we examine point and interval forecast accuracies in Section~\ref{sec: results}. Conclusions are presented in Section~\ref{sec: conclusion}, along with some ideas on how the present research can be extended.

\section{Intraday CBOE volatility index}\label{sec: VIX}

\begin{figure}[!htbp]
\centering
\subfloat[Functional time series curves of intraday VIX index]
{\includegraphics[width=8cm]{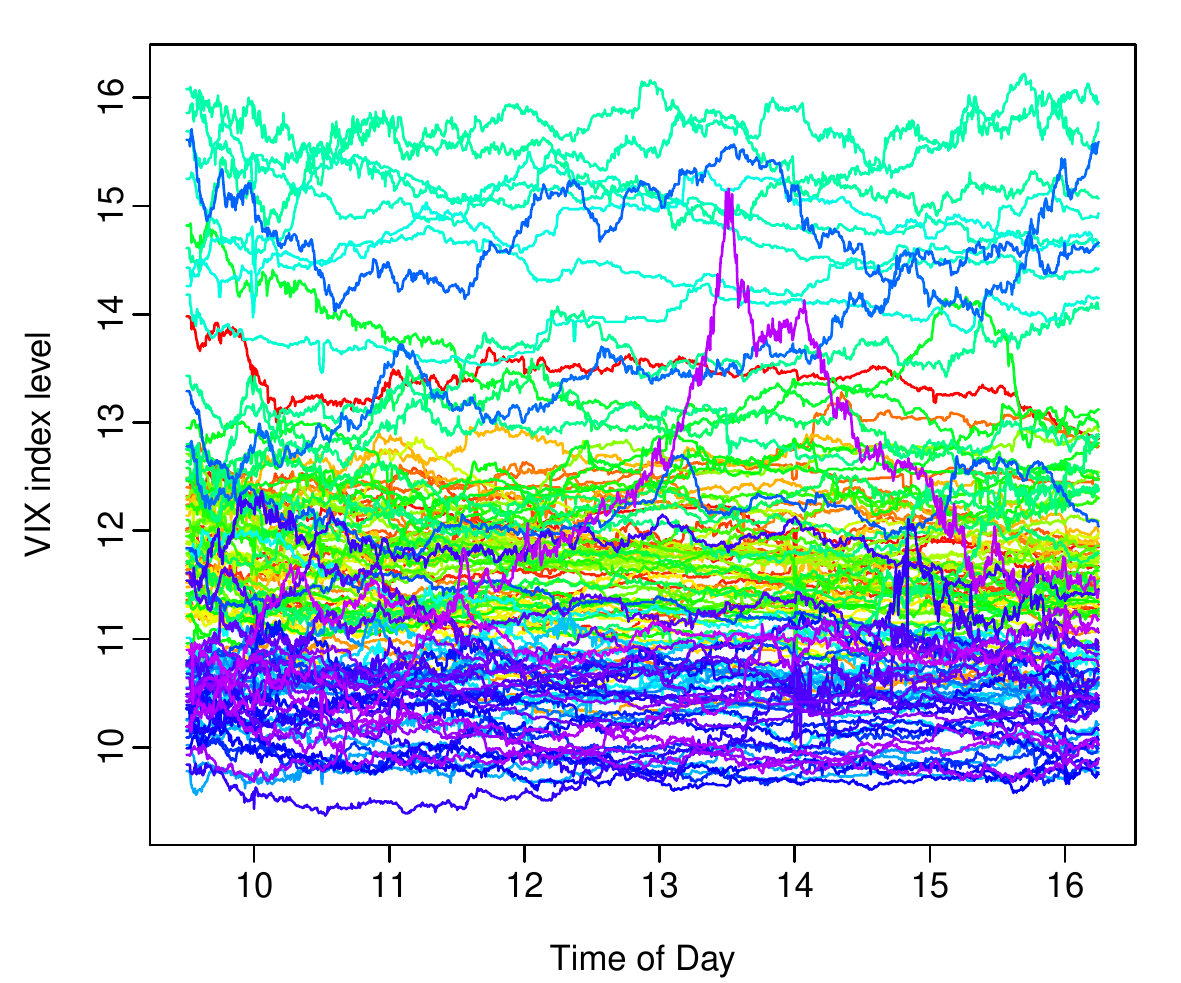}}
\qquad
\subfloat[Functional time series curves of CIDR VIX index]
{\includegraphics[width=8cm]{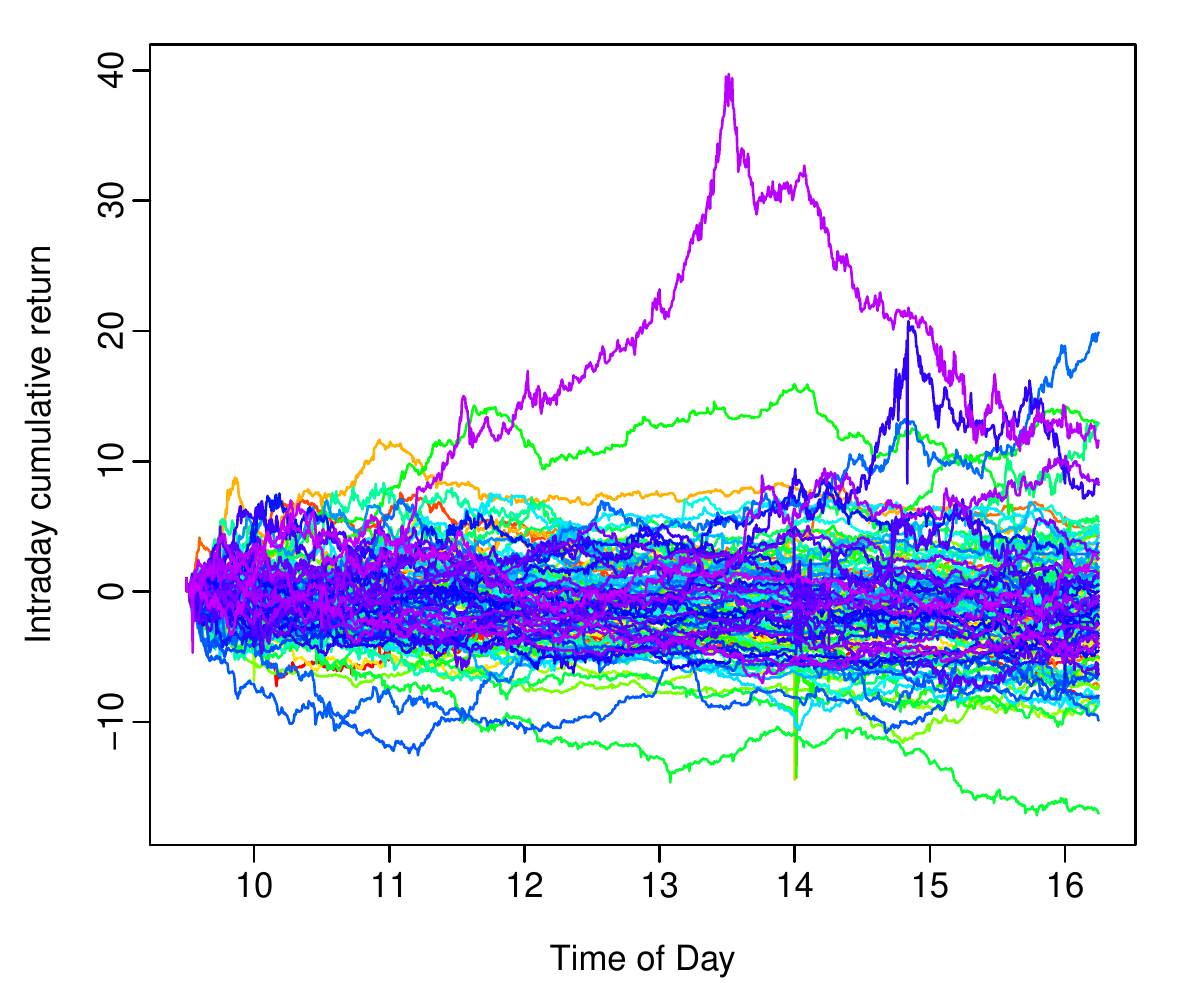}}
\caption{Graphical displays of VIX index functional data curves from 3 January 2017 to 30 June 2017.}\label{fig: 1}
\end{figure}

As the CBOE calculates the VIX index every 15 seconds\footnote{The use of 15-second frequency is because it is the highest frequency available for the VIX index. Some of its derivatives may trade at a higher frequency, but the index value is only recalculated and released on a 15-second basis. So to provide us with the most current information and the highest level of granularity to model its evolution, we use these 15-second VIX quotes. This data frequency is particularly relevant for the dynamic functional time series approaches we propose.}, we adopt the highest resolution dataset and consider 15-second VIX index values from 3 January 2017 to 30 June 2017. Via the functional time series analysis techniques, non-stationary daily curves of intraday VIX values are transformed into daily curves of cumulative intraday returns \citep[CIDRs, see, e.g.,][]{GRH10}. CIDRs have been proved to be stationary by \cite{KMZ14}. Although the VIX itself is not directly tradable, the approach is employed as it offers a transformation that facilities the construction of stationary functional time series. Let $P_i(t_j)$ denote the daily VIX value at time $t_j$ ($j = 1, \cdots, m$) on day $i$ ($i = 1,\cdots, n$). CIDRs are computed as 
\begin{equation}
R_i(t_j) = 100 \times \left[ \ln P_i(t_j) - \ln P_i(t_1) \right],\label{eq 1}
\end{equation}
where $\ln(\cdot)$ represents the natural logarithm and time intervals between $t_{j-1}$ and $t_j$ are 15 seconds in length. Figure~\ref{fig: 1} indicates that the curved shape of CIDRs is similar to the shape of the VIX curves at the original scale. 

Thus, we consider constructing continuous curves of CIDRs on the observed VIX index values as 
\begin{equation*}
\mathcal{X}_i(t) = R_i(t_j), \qquad t \in \left( 15(j-1), 15j \right], \qquad \text{for} \quad  j = 1, \cdots, \tau, 
\end{equation*}
where $\tau$ denotes the total number of realised 15-second intraday cumulative returns. In this study, we consider VIX values reported by CBOE between 9:30:00 to 16:15:00 on each trading day. Therefore, $\tau$ has a maximal value of 1621. Once the functional time series is constructed, we work directly with this continuous series going forward.

To construct continuous CIDRs, we initially group 201,500 discrete observations obtained between 9:30:00 to 16:15:00 into 125 daily curves. We observe that on different trading days, timings of the first and the last VIX values can vary slightly. To ensure each daily curve is of the same length, we adopt the linear interpolation algorithm \citep{Hyndman17} in \verb|R| software \citep{R} to fill in missing values (if any) at both ends of each daily curve. CIDRs are then computed based on~Equation~\eqref{eq 1}, leading to $N = 202,625 $ discrete intraday cumulative returns separated into $n = 125$ CIDRs.

\begin{figure}[!htbp]
\centering
\subfloat[Bivariate HDR boxplot]
{{\includegraphics[width=8cm]{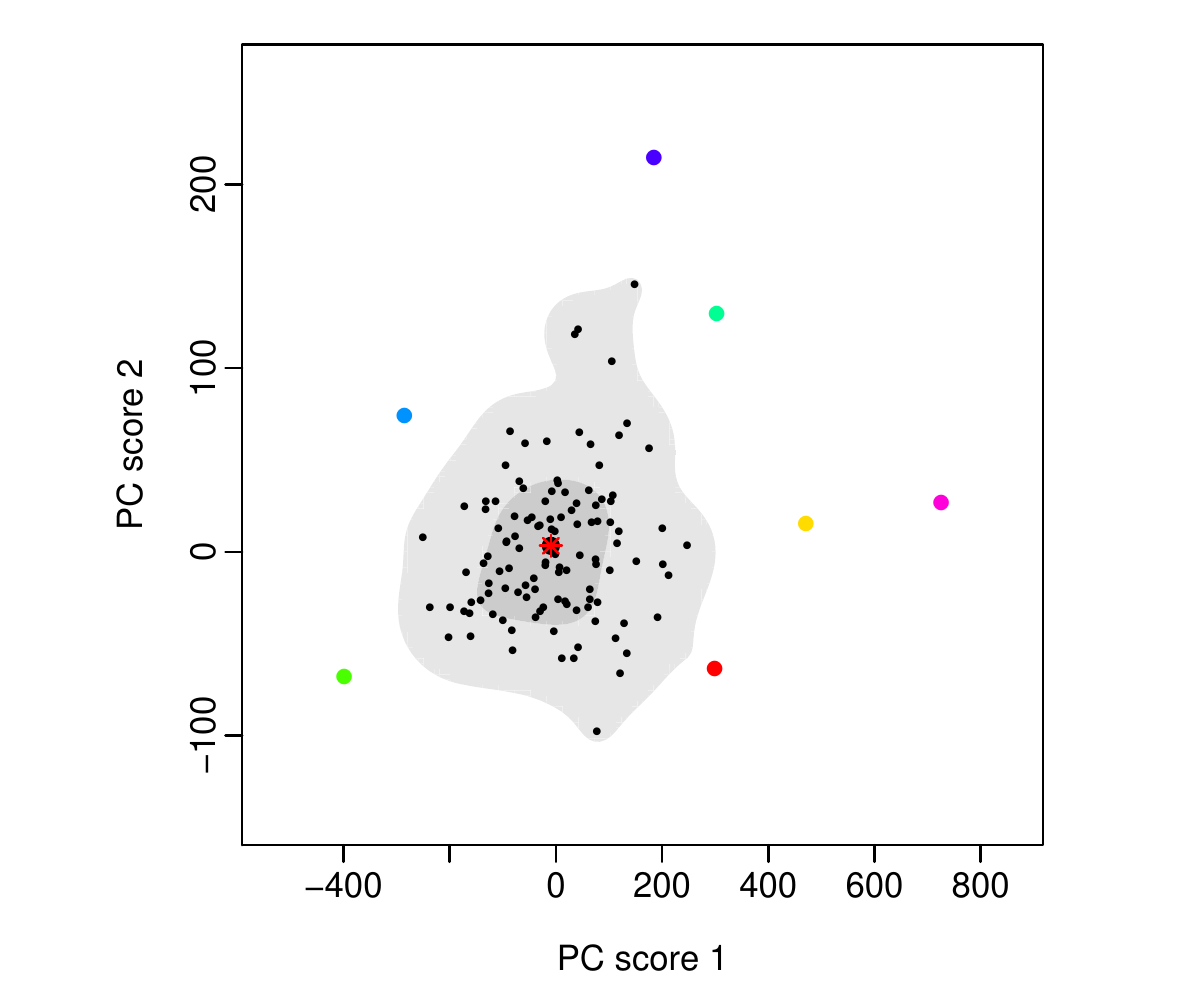} }\label{fig:2a}}
\qquad
\subfloat[Functional HDR boxplot (mode is shown as the black line)]
{{\includegraphics[width=8cm]{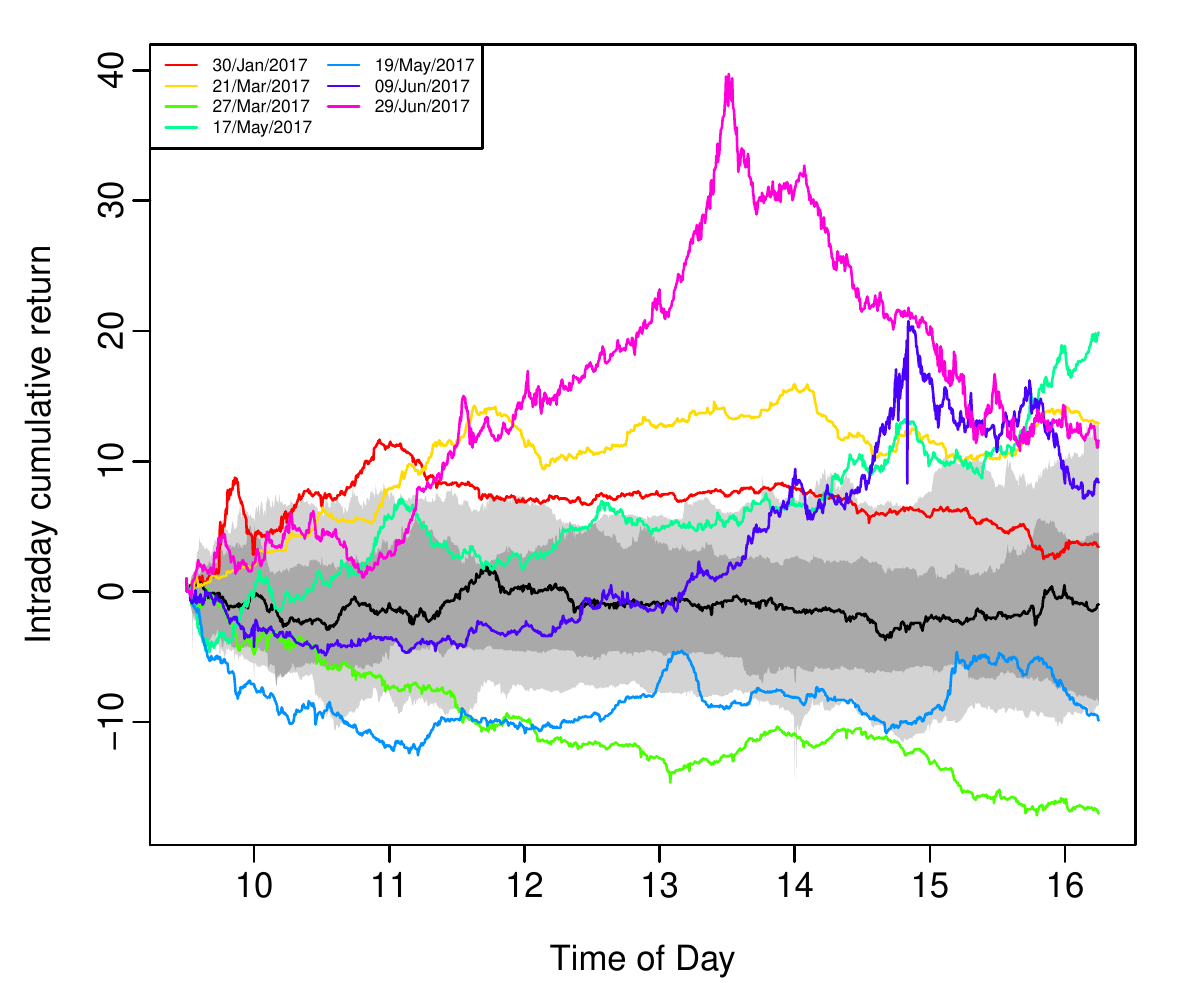} }\label{fig:2b}}
\caption{A functional outlier detection method, namely the highest density region, is adapted to identify five outliers representing about 5\% of the total number of curves.}\label{fig: 2}
\end{figure}

Traditionally, the presence of outliers can diminish the performance of functional time series models. Therefore, we showcase that our approach is robust to the presence of such outliers. According to \cite{FGG07} and \cite{HS10}, a functional outlier is a curve generated by a stochastic process that has a different distribution from that of standard curves. Based on this general definition, outliers can be further grouped into categories, such as magnitude outlier, shape outlier, and partial outlier \citep[see][for more details]{SPL16}. Using the functional outlier detection method proposed by \cite{HS10}, among the 125 curves the seven most volatile daily curves (about 5\%)\footnote{The proportion of outlying curves is specified to be around 5\% which is an arbitrary but commonly used threshold in functional data analysis.} are suspected to be outliers, as shown in Figure~\ref{fig: 2}. Identified outliers that display abnormal shapes and magnitudes correspond to the dates 30 January 2017, 21 March 2017, 27 March 2017, 17 May 2017, 19 May 2017, 9 June 2017 and 29 June 2017 (as highlighted by the coloured lines in Figure~\ref{fig:2b}). These days correspond with events of special economic significance as we now outline. The first highlighted trading day of 30 January 2017 witnessed a stock market hit by US travel ban fears as President Trump placed a ban on immigration from seven Muslim-majority countries. The most volatile curve displayed corresponded to 29 June 2017 when technology stocks of S\&P 500 slumped suddenly amid concerns of overvaluation and monetary policy tightening from the Federal Reserve causing investors to dump\footnote{A comprehensive discussion of these special economic events leading to unusual market movements are included in our Supplement.}.

We consider functional time series forecasting methods that are robust to the effect of outliers. To exhibit this desirable property, we have not excluded any of those seven potentially outlying curves mentioned in this section from either our parameter estimation or forecasting processes.

\section{Forecasting method}\label{sec: methodology}

We employ the functional principal component regression of \cite{HS09,Shen09,HSH12} and \cite{ANH15} to model and forecast functional time series of CIDRs. We adopt the functional PCA as the primary forecasting technique because it plays a vital role in the development of functional data analysis \citep[for reviews, see][]{Hall11, Morris15, Shang14}. To deal with outliers identified, we consider a robust functional PCA method \citep{HU07} and a robust regularised singular value decomposition \citep{ZSH13}.

\subsection{Functional principal component regression}\label{subsec fpca}

We denote an arbitrary stationary functional time series by $\left(\mathcal{X}_i: i \in \mathbb{Z} \right)$. $\mathcal{X}_i$ are assumed to be located in the Hilbert space, $\mathcal{H} = L^2(\mathcal{I})$ which has the inner product $\langle x, y\rangle = \int_{\mathcal{I}} x(t) y(t) dt $, where $t$ is a continuum and $\mathcal{I} \in [1,p]$ represents a function support range. $\mathcal{X}_i$ are assumed to be square integrable with finite squared norm $\norm{\mathcal{X}_i}^2 = \int_{\mathcal{I}} \mathcal{X}_i^2 (t) dt  < \infty $. Using $\bm{\mathcal{X}}$ to denote a set of all $\mathcal{X}_i$ defined on a common probability space $(\varOmega, A, P)$, we have $\bm{\mathcal{X}} \in L^p_H (\varOmega, A, P) $, indicating that for some positive value $p >0$, $\E(\norm{\mathcal{X}}^p) < \infty$. When $p=1$, it gives the mean curve of the stationary functional time series $\mathcal{X}(t)$, defined as $\mu(t)  = \E [\mathcal{X}(t)] $; when $p =2$, it gives the covariance operator $\mathcal{K}(s,t) = \cov[\X(s),\X(t) ] = \E \{ [ \X(s) - \mu(s)][ \X(t) - \mu(t)]   \} $. According to \cite{Karhunen1946} and \cite{Loeve1946}, the covariance operator can be defined as
\begin{equation*}
\mathcal{K} (s,t) = \sum_{k=1}^{\infty} \lambda_{k} \phi_{k} (s) \phi_{k}(t), \qquad s,t\in\mathcal{I},
\end{equation*}
where $\phi_{k}(t)$ denotes the $k$\textsuperscript{th} orthonormal principal component, and $\lambda_k$ denotes the $k$\textsuperscript{th} eigenvalue. 

Separability of the Hilbert space allows us to apply the Karhunen-Lo\`{e}ve expansion to approximate the stochastic process $\X$ based on the principal component score $\beta_{k}$ given by the projection of $[\X(t) - \mu (t)]$ in the direction of the $k$\textsuperscript{th} eigenfunction $\phi_k$ as
\begin{equation}
\X(t) = \mu(t) + \sum_{k=1}^{\infty} \beta_k \phi_k(t). \label{eq kl}
\end{equation}
Dimension reduction can be achieved by utilising Equation~\eqref{eq kl} and truncating the first $K$ principal components. In this way, we assume that the $K$-dimensional vector $(\beta_1, \cdots, \beta_k)$ contains most of the useful information in $\X(t)$, leading to an adequate approximation as 
\begin{equation}
\X(t) = \mu(t) + \sum_{k=1}^{K}\beta_k\phi_k(t) + e(t), \label{eq xt}
\end{equation}
where $e(t)$ represents model residuals after excluding the first $K$ principal components. There are at least four approaches to selecting the number of retained principal components $K$: 
\begin{inparaenum}
\item[(1)] the scree plot or the fraction of variance explained by the first several functional principal components \citep{Chiou12}; 
\item[(2)] the pseudo-versions of the Akaike information criterion (AIC) and Bayesian information criterion \citep{YMW05}; 
\item[(3)] the cross-validation with one-curve-leave-out \citep{RS91}; 
\item[(4)] the bootstrap technique \citep{HV06}.  
\end{inparaenum}
We use the first method to find the value of $K$ satisfying the following condition:
\begin{equation*}
K = \argmin_{K:K\geq1} \left\lbrace \sum_{k=1}^{K} \widehat{\lambda}_k \Big/ \sum_{k=1}^{\infty} \widehat{\lambda}_k \mathds{1}_{\left\{\widehat{\lambda}_k > 0\right\}} \geq \delta  \right\rbrace, 
\end{equation*}
where $\delta = 90\%$ and $\mathds{1}_{\left\{\widehat{\lambda}_k > 0\right\}} $ to exclude possible zero eigenvalues\footnote{\cite{HK12} compared different methods of selecting the number of retained functional principal components and argued that the cumulative percentage of total variance method worked best throughout applications in their book. Hence, we have adopted this method in our study of VIX.}.

In practice, construction of a functional time series $\bm{\X(t)} = \left\{ \X_1(t) , \cdots, \X_n(t) \right\} $ utilises the empirical mean function $\widehat{\mu}(t) = \frac{1}{n}\sum_{i=1}^{n} \X_i (t) $ and the empirical covariance function $\widehat{\mathcal{K}}(s,t)$ estimated based on data. Estimated functional principal components functions $\bm{\varPhi}(t) = \left[\widehat{\phi}_1(t), \cdots, \widehat{\phi}_k(t)\right] $ are then extracted from the empirical covariance function. The $h$-step-ahead point forecasts of curve $\X_{n+h}(t)$ can then be obtained through a conditional expectation as
\begin{align*}
\widehat{\X}_{n+h|n}(t) &= \E\left[ \X_{n+h}(t)| \widehat{\mu}_t,  \bm{\varPhi}(t), \bm{\X}(t) \right] \\
&= \widehat{\mu}(t) + \sum_{k=1}^{K} \widehat{\beta}_{n+h|n,k} \widehat{\phi}_k(t),
\end{align*}
where $\widehat{\beta}_{n+h|n,k}$ represents the point forecasts of $\beta_{n+h,k}$. A univariate time series forecasting method and a multivariate time series forecasting method are used in the analysis of our VIX data and discussed in Section~\ref{subsec forecasting}.

\subsection{Robust functional principal component analysis}\label{subsec rpca}

The presence of outliers often affects the estimation of functional principal components from the covariance operator $\mathcal{K}(s,t) = \cov[\X(s), \X(t)]$ and leads to inferior estimation and forecast accuracies. \cite{HU07} introduced a two-step algorithm of extracting robust functional principal components by down-weighting the effect of outliers. This procedure begins by initially implementing the robust functional PCA algorithm as proposed by \cite{HRV02}, calculating the integrated squared error of curve $i$ as 
\begin{equation*}
v_i = \int_{ t \in \mathcal{I}} \Big[  \X_i(t) - \sum_{k=1}^{K} \beta_{i,k}\phi_{k} (t) \Big] ^2 dt, \qquad \text{for} \quad i=1,\cdots,n.
\end{equation*}
By this algorithm, outlying curves tend to have large values of $v_i$. A set of weights can be assigned to curves satisfying $v_i < s + \lambda \sqrt{s}$, where a tuning parameter $\lambda$ controls the amount of robustness and $s$ represents the median of $\{v_1, 
\cdots, v_n  \}$. According to \cite{HU07}, the efficiency of this procedure follows a cumulative normal distribution. We specify $\lambda = 2.33$ to ensure the efficiency equals to $\varPhi\left(2.33/\sqrt{2}\right) = 95\%$, suggesting that 5\% of curves are to be classified as outliers.

\subsection{Robust regularised singular value decomposition}\label{subsec rsvd}

Apart from conducting robust PCA, we also consider the robust regularised singular value decomposition (RobRSVD) method of \cite{ZSH13} for analysing functional data. RobRSVD extends the idea of the traditional singular value decomposition (SVD), which aims to find a sequence of rank-one matrix approximations of a data matrix \citep{GZ79}, to obtain a series of robust regularised rank-one matrix approximations. Implementation of the RobRSVD method involves computing the first pair of singular vectors via solving a least-squares problem as
\begin{equation*}
(\bm{u}_1, \bm{v}_1) = \argmin_{\bm{u},\bm{v}} \left\lbrace \rho \left(\bm{\mathcal{X}} -  \bm{u} \bm{v}^{\top}\right)  + \mathcal{P}_{\bm{\lambda}}(\bm{u},\bm{v}) \right\rbrace,
\end{equation*}
where $\bm{u}$ and $\bm{v}$ are $m$-dimensional and $n$-dimensional vectors respectively, $\rho (\cdot)$ is a robust loss function, $\mathcal{P}_{\bm{\lambda}}(\bm{u},\bm{v})$ is a two-way roughness penalty to ensure smoothness for the $\bm{u}$ and $\bm{v}$, and $\bm{\lambda}$ is a vector of penalty parameters. After removing the effects of the first pairs of $\bm{u}$ and $\bm{v}$, following pairs of vectors can be obtained by applying the method to the model residuals from the previous step. Features of this sequential approach include allowing the different pairs of singular vectors to have different levels of smoothness. According to \cite{ZSH13}, the nonnegative, symmetric loss function for a rank-one approximation of the matrix $\bm{\mathcal{X}}$ is defined as
\begin{equation}
\rho\left( \frac{\bm{\mathcal{X}} -  \bm{u} \bm{v}^{\top}}{\sigma}   \right)  = \sum_{i=1}^{m} \sum_{j=1}^{n} \rho \left( \frac{\mathcal{X}_{i}(t_j) - u_i v_j}{\sigma}   \right),\label{eq 3}
\end{equation}
where $\sigma$ is a scale parameter measuring the variability in the approximation error. In practice, the value of $\sigma$ can be estimated by the conventional normalised median absolute deviation method, as defined by \cite{MM06} as
\begin{equation*}
\widehat{\sigma} = \frac{1}{0.675}\text{Med}_{ij} \left( |r_{ij}|, r_{ij} \neq 0 \right), 
\end{equation*}
where residuals are defined as $r_{ij} = \bm{\mathcal{X}} - \widehat{\bm{u}} \widehat{\bm{v}}^{\top}$. \cite{HSB09} suggested a specific form of the penalty function
\begin{equation*}
\mathcal{P}_{\bm{\lambda}}(\bm{u},\bm{v}) = \lambda_{\bm{u}} \bm{u}^{\top} \bm{\varOmega}_{\bm{u}} \bm{u} \cdot \norm{\bm{v}}^2 + \lambda_{\bm{v}} \bm{v}^{\top} \bm{\varOmega}_{\bm{v}} \bm{v} \cdot \norm{\bm{u}}^2 + \lambda_{\bm{u}} \bm{u}^{\top} \bm{\varOmega}_{\bm{u}} \bm{u} \cdot \lambda_{\bm{v}} \bm{v}^{\top} \bm{\varOmega}_{\bm{v}} \bm{v},
\end{equation*}
where $\bm{\varOmega}_{\bm{u}}$ and $\bm{\varOmega}_{\bm{v}}$ are symmetric and nonnegative definite penalty matrices to the left and right singular vectors respectively, and $\norm{\cdot}$ is the Euclidean norm. The implementation of Equation~\eqref{eq 3} includes a Huber's function defined as
\begin{equation*}
\rho_{\theta}(x) = 
\begin{cases}
x^2 & \text{if $|x| \leq \theta$} \\
2\theta|x| - \theta^2 & \text{if $|x| > \theta$}
\end{cases},
\end{equation*}
where a smaller value of the smoothness parameter $\theta$ gives a more robust estimation. A commonly used $\theta = 1.345$ is adopted because it produces 95\% efficiency for normal errors \citep{HR09}.

\subsection{Univariate and multivariate time series forecasting methods}\label{subsec forecasting}

We consider the univariate time series forecasting method of \cite{HS09} which utilises an ARIMA model. This procedure can be applied to non-stationary time series that contains a stochastic trend component. The VIX data considered spans 125 days without any apparent seasonality. Therefore, an ARIMA model is defined in the following general form:
\begin{align*}
\left(1 - \phi_1\mathcal{B} - \cdots - \phi_p\mathcal{B}^p\right)\left(1-\mathcal{B}\right)^d \bm{\beta}_k = \gamma + \left(1 + \theta_1 \mathcal{B} + \cdots + \theta_q \mathcal{B}^q\right) \bm{w}_k,
\end{align*}
where $\gamma$ represents the intercept, $(\phi_1, \cdots, \phi_p)$ denote the coefficients associated with the autoregressive component, $(\theta_1, \cdots, \theta_q)$ denote the coefficients associated with the moving average component, $\mathcal{B}$ denotes the backshift operator, $d$ denotes the differencing operator, $\bm{\beta}_k = (\beta_{1,k}, \cdots, \beta_{n,k})$ represents the $k$\textsuperscript{th} estimated principal component scores, and $\bm{w}_k = \left(w_{1,k}, \cdots, w_{n,k}\right)$ represents the error term. We select the optimal ARIMA model based on an information criterion, and subsequently estimate parameters in the optimal model via the maximum likelihood estimation method. In practice, we use the automatic ARIMA algorithm of \cite{HK07} to select the optimal model based on the corrected AIC (an information criterion that performs well for small sample sizes).

For data that have strict diagonal auto-covariances at all lags, univariate time series forecasting methods can be accurate and time efficient \citep[see, e.g.][]{ANH15}. Given the vectors of functional principal component scores may have the dependence structure of the principal component score matrix, a univariate functional time series modelling process may result in a loss of information. \cite{ANH15} introduced a multivariate time series model to counter this problem. The most commonly used multivariate time series model is the vector autoregressive (VAR) model, for the following reasons: 
\begin{inparaenum}[(1)]
\item estimation of parameters for the VAR model is generally straightforward as ordinary least squares (OLS), maximum likelihood and Bayesian methods are all possible candidate procedures;
\item the properties of the VAR model have been studied \citep[see, e.g. ][]{Tsay13};
\item the VAR model has a clear structure and can be considered as a type of multivariate multiple linear regression.
\end{inparaenum}
To utilise the multivariate time series forecasting method, we define a VAR model of order $\vartheta$ if the multivariate principal component scores $\bm{\beta}_k = [\beta_{1,k}, \cdots, \beta_{n,k}]^{\top}$ satisfy
\begin{equation*}
\bm{\beta_k = \phi_0 + \sum_{v=1}^{\vartheta} \phi_{\vartheta} \beta_{k-v} + a_k },
\end{equation*}
where $\bm{\phi}_0$ is an $n$-dimensional constant vector, and $\bm{\phi}_v$ are $n \times n$ matrices for $v > 0$ and $\bm{\phi}_{\vartheta} \neq 0$ and $\bm{a}_k$ is a set of independent and identically distributed (i.i.d.) random error vectors with a mean of zero and a positive-definite covariance matrix that has only positive eigenvalues. We use the corrected AIC to select the optimal order of $\vartheta$.

Comparisons of point and interval forecast accuracies between the univariate ARIMA and multivariate VAR models can be found in the literature. For example, \cite{PS07} noted that if the series (i.e., principal component score vectors) are very weakly related, considering the joint dynamics of the series can marginally improve the univariate time series forecasts. Conversely, when historical observations of one time series can significantly influence another time series, the multivariate time series forecasts have many advantages compared to the univariate time series forecasts. We present a comparison between point and interval forecasts obtained by the univariate and multivariate time series forecasting methods in Section~\ref{sec: results}.

\section{Updating point forecasts}\label{sec: updating}

Constructing functional time series based on segments of a longer univariate time series often encounters the problem that newly arriving curve data may not possess all the important features of the complete curve. After observing the first $m_0$ time periods on day $n+1$ of $\X_{n+1}(t)$, denoted by $\X_{n+1}(t_e) = [ \X_{n+1}(t_1), \cdots, \X_{n+1}(t_{m_0}) ]^{\top}$, our interest is forecasting the data in the remaining part of the same curve, denoted by $\X_{n+1}^l(t)$, where $t \in \mathcal{I}_l$. As described in Section~\ref{sec: methodology}, the one-step-ahead TS forecast of 
\begin{equation}
\widehat{\X}_{n+1|n}^{l,\text{TS}} (t) = \widehat{\mu}^l (t) + \sum_{k=1}^{K} \widehat{\beta}^{\text{TS}}_{n+1|n,k} \widehat{\phi}_k^l(t) , \qquad \text{for} \quad l \in (m_0,p], \label{eq ts forecast}
\end{equation}
where $\widehat{\mu}^l (t)$ represents the mean curve for the remaining time period and $\widehat{\phi}_k^l (t)$ represents the $k$\textsuperscript{th} functional principal component of the remaining time period.

However, Equation~\eqref{eq ts forecast} does not utilise the partially observed curve. For improving point forecast accuracy, dynamic updating methods are often considered because such methods incorporate newly observed information into forecasting the remaining curve of the day $n+1$ \citep[see also][]{SH01}. In this section, several dynamic updating methods are introduced and compared in consideration of their ability to improve forecast accuracy.

\subsection{Block moving (BM)}\label{subsec bm}

The BM method adopts the general structure of the TS method with redefined starting and ending time points for our CIDR curves. Treating time as a continuous variable, the BM method changes the support range of Equation~\eqref{eq xt} from $[1,p]$ to $(m_0, p] \cup [1, m_0] $. The consequences include a loss of data at the beginning of the first curve and complete subsequent partially observed curves. The data loss in the first curve will only have a minimal negative influence on the forecasts given enough data (i.e., a large number of curves) because the forecasts depend very little on observations from the distant past.

\subsection{Ordinary least squares (OLS)}\label{sec ols}

A regression approach based on the functional principal components obtained in Equation~\eqref{eq xt} can also be used to approximate the remaining part of the most recent curve. Let $\bm{\mathcal{F}}_e$ be the $m_0 \times K$ matrix of the response variable, whose $(j,k)$\textsuperscript{th} entry is $\widehat{\phi}_k(t_j)$, for $1\leq j \leq m_0$ and $1 \leq k \leq K$. Let $\bm{X}_{n+1}(t_e)$ be the $m_0 \times 1$ matrix of the response variable, $\bm{\beta}_{n+1} = [\beta_{n+1,1}, \cdots, \beta_{n+1,K}]^{\top}$ and $\bm{\varepsilon}_{n+1}(t_e) = [\varepsilon_{n+1}(t_1), \cdots, \varepsilon_{n+1}(t_{m_0})]^{\top}$. The regression model used to estimate coefficients $\bm{\beta}_{n+1}$ has the following equation
\begin{equation*}
\bm{X}^{*}_{n+1}(t_e) = \bm{\mathcal{F}}_e \bm{\beta}_{n+1} + \bm{\varepsilon}_{n+1}(t_e),
\end{equation*}
where $\bm{X}^{*}_{n+1}(t_e) = \bm{X}_{n+1}(t_e) - \bm{\widehat{\mu}}(t_e)$ denotes the mean-adjusted response variable. Using the OLS estimation method, we have
\begin{equation}
\bm{\widehat{\beta}}_{n+1}^{\text{OLS}} = (\bm{\mathcal{F}}^{\top}_e \bm{\mathcal{F}}_e)^{-1} \bm{\mathcal{F}}_e^{\top} \bm{X^{*}}_{n+1}(t_e). \label{eq ols}
\end{equation}
Then, the OLS forecasts of $\X_{n+1}^l(t)$ at discretised time points are given by
\begin{equation*}
\bm{\widehat{X}}_{n+1}^{\text{OLS}}(t_l) = \widehat{\mu}(t_l) + \sum_{k=1}^{K} \bm{\widehat{\beta}}_{n+1,k}^{\text{OLS}} \widehat{\phi}_k(t_l),
\end{equation*}
where $t_l = \{ t_{m_0 +1}, \cdots, t_{\tau};   \tau \leq q  \}$ represents the discretised time points in the remaining period\footnote{When updating point forecasts for a particular day, we keep using the same $K$ obtained in the model estimation step as per Equation~\eqref{eq xt}.}.

\subsection{Ridge regression (RR)}\label{subsec rr}{}

To conduct the inverse computation in Equation~\eqref{eq ols} in practice, a sufficiently large number of observations for $\bm{\widehat{\beta}}_{n+1}^{\text{OLS}} = \left[ \widehat{\beta}_{1,n+1}^{\text{OLS}}, \cdots,  \widehat{\beta}_{K,n+1}^{\text{OLS}}  \right]^{\top} $ are necessary. To address this problem, we consider the ridge regression method of \cite{HK70}, using the functional principal components as predictors and the partially observed data as responses. The RR method has a feature of shrinking the regression coefficient estimates towards zero, achieved by minimising a penalised residual sum of squares:
\begin{equation*}
\argmin \left\lbrace  \left(\bm{X}^{*}_{n+1}(t_e) - \bm{\mathcal{F}}_e \bm{\beta}_{n+1}\right)^{\top}   (\bm{X}^{*}_{n+1}(t_e) - \bm{\mathcal{F}}_e \bm{\beta}_{n+1}) + \lambda \bm{\beta}_{n+1}^{\top} \bm{\beta}_{n+1}    \right\rbrace,
\end{equation*}
where $\lambda > 0$ is a parameter that controls the amount of shrinkage. Taking the first derivative with respect to $\bm{\beta}_{n+1}$ leads to the estimated coefficients as
\begin{equation*}
\widehat{\bm{\beta}}_{n+1}^{\text{RR}} = \left(\bm{\mathcal{F}}_e^{\top} \bm{\mathcal{F}}_e + \lambda \bm{I_K}   \right)^{-1} \bm{\mathcal{F}}_e^{\top}  \bm{X}^{*}_{n+1}(t_e),
\end{equation*}
where $\bm{I}_K$ is a $(K \times K)$ identity matrix. The $\widehat{\bm{\beta}}_{n+1}^{\text{RR}}$ has the following properties:
\begin{enumerate*}[label=(\roman*)]
\item when $0<\lambda<\infty$, $\widehat{\bm{\beta}}_{n+1}^{\text{RR}} $ is a weighted average between 0 and $\widehat{\bm{\beta}}_{n+1}^{\text{OLS}}$;
\item when $\lambda \rightarrow 0$, $\widehat{\bm{\beta}}_{n+1}^{\text{RR}} $ approaches to $\widehat{\bm{\beta}}_{n+1}^{\text{OLS}} $;
\item when $\lambda \rightarrow \infty$, $\widehat{\bm{\beta}}_{n+1}^{\text{RR}} $ goes to 0 .
\end{enumerate*}
With an optimal selection of $\lambda$, the RR forecasts of $\X_{n+1}^l (t)$ at discretised time points are given by
\begin{equation*}
\widehat{\bm{X}}_{n+1}^{\text{RR}}(t_l) = \widehat{\mu}(t_l) + \sum_{k=1}^{K}\widehat{\beta}_{n+1,k}^{\text{RR}}\widehat{\phi}_k(t_l).
\end{equation*}

\subsection{Penalised least squares (PLS)}\label{subsec pls}

The RR method avoids the singularity problem associated with $\widehat{\bm{\beta}}_{n+1}^{\text{OLS}}$, but it does not use the TS forecasts $\widehat{\bm{\beta}}_{n+1|n}^{\text{TS}}$. The PLS method shrinks regression coefficient estimates towards 
$\widehat{\bm{\beta}}_{n+1|n}^{\text{TS}}$ by minimising a penalised residual sum of squares:
\begin{equation}
\argmin \left\lbrace  (\bm{X}^{*}_{n+1}(t_e) - \bm{\mathcal{F}}_e \bm{\beta}_{n+1})^{\top}   (\bm{X}^{*}_{n+1}(t_e) - \bm{\mathcal{F}}_e \bm{\beta}_{n+1}) + \lambda ( \bm{\beta}_{n+1} - \widehat{\bm{\beta}}_{n+1|n}^{\text{TS}})^{\top} (\bm{\beta}_{n+1} - \widehat{\bm{\beta}}_{n+1|n}^{\text{TS}})   \right\rbrace. \label{eq pls argmin}
\end{equation}

In Equation~\eqref{eq pls argmin}, the first term measures the ``goodness of fit" and the second term penalises the departure of the regression coefficient estimates from the $\widehat{\bm{\beta}}_{n+1|n}^{\text{TS}}$. Taking the first derivative with respect to $\bm{\beta}_{n+1}$ leads to the estimated coefficients as 
\begin{equation}
\widehat{\bm{\beta}}_{n+1}^{\text{PLS}} = \left(\bm{\mathcal{F}}_e^{\top} \bm{\mathcal{F}}_e + \bm{\lambda I_K}   \right)^{-1}  \left( \bm{\mathcal{F}}_e^{\top} \bm{X}^{*}_{n+1}(t_e) + \lambda \widehat{\bm{\beta}}_{n+1|n}^{\text{TS}}  \right). \label{eq beta estimate}
\end{equation}
The $\widehat{\bm{\beta}}_{n+1}^{\text{PLS}}$ has the following properties:
\begin{enumerate*}[label=(\roman*)]
\item when $0<\lambda<\infty$, $\widehat{\bm{\beta}}_{n+1}^{\text{PLS}} $ is a weighted average between $\widehat{\bm{\beta}}_{n+1|n}^{\text{PLS}}$ and $\widehat{\bm{\beta}}_{n+1}^{\text{OLS}}$;
\item when $\lambda \rightarrow 0$, $\widehat{\bm{\beta}}_{n+1}^{\text{PLS}} $ approaches to $\widehat{\bm{\beta}}_{n+1}^{\text{OLS}} $ given $\left(\bm{\mathcal{F}}_e^{\top} \bm{\mathcal{F}}_e  \right)^{-1}$ exists;
\item when $\lambda \rightarrow \infty$, $\widehat{\bm{\beta}}_{n+1}^{\text{PLS}} $ go to $\widehat{\bm{\beta}}_{n+1}^{\text{OLS}}$.
\end{enumerate*}
With an optimal selection of $\lambda$, the PLS forecasts of $\X_{n+1}^l (t)$ at discretised time points are given by
\begin{equation*}
\widehat{\bm{X}}_{n+1}^{\text{PLS}}(t_l) = \widehat{\mu}(t_l) + \sum_{k=1}^{K}\widehat{\beta}_{n+1,k}^{\text{PLS}}\widehat{\phi}_k(t_l).
\end{equation*}

\subsection{Selection of shrinkage parameters in the RR and PLS methods}\label{subsec select lambda}

Both the RR method and the PLS method require the optimal selection of $\lambda$. To minimise forecast errors, the choice of $\lambda$ should also vary as we observe an increasing number of data points in a trading day. We adopt a holdout forecast evaluation method to select $\lambda$. We divide a time series of functions into a training sample (including days from 1 to 85) and a testing sample (including days from 86 to 125). We further divide the training sample into a training set (including days from 1 to 43) and a validation set (including days from 44 to 85). The optimal values of $\lambda$ for different updating periods are determined by minimising the averaged forecast error criteria within the validation set (see Section~\ref{sec: accuracy}). 

\subsection{Functional linear regression}\label{subsec flr}

The OLS, RR, and PLS methods share a common feature of considering a discretised data approach to produce updates for the remaining period. To take advantage of functional data analysis, which allows observations to be considered as realisations of a continuous function, we also consider the functional linear regression of \cite{Chiou12} \citep[see also][]{MSS11}. For a detailed overview of functional linear regression, refer to \cite{Muller05} and \cite{Morris15}. The functional linear regression can be expressed as 
\begin{equation}
\X_{n+1}^l (t) = \mu^l(t) + \int_{ s \in \mathcal{I}_e} \left[\X_{n+1}^e(s) - \mu^e(s) \right] \beta(s,t) \text{ds} + e^l_{n+1}(t) \qquad s \in  \mathcal{I}_e, \ t \in  \mathcal{I}_l, \label{eq flr}
\end{equation}
where $\mathcal{I}_e \in [1,m_0]$ and $\mathcal{I}_l \in (m_0,p]$ represent two function support ranges for the partially observed and remaining segments of curve on day $n+1$, respectively; $\mu^e(s)$ and $\mu^l(t)$ represent two mean functions for the partially observed data and remaining data periods, respectively; $\X_{n+1}^e(s)$ and $\X_{n+1}^l(t)$ represent functional predictor and functional response variables, respectively. Equation~\eqref{eq flr} can be viewed as function-on-function linear regression \citep[see also][Chapter 16]{RS06}, where $\beta(s,t)$ and $e^l_{n+1}(t)$ denote the regression coefficient function and error function, respectively. 

Estimation of the regression coefficient function $\beta(s,t)$ involves projecting a time series of functions onto functional principal component scores. Using functional PCA, we obtain
\noindent

\begin{minipage}{.5\linewidth}
\begin{align*}
\X_i^e(t) &= \mu^e(t) + \sum_{k=1}^{\infty} \xi_{i,k}\phi_k^e(t) \\
&= \mu^e(t) + \sum_{k=1}^{K}\xi_{i,k}\phi_k^e(t) + \eta_i^e(t), \qquad
\end{align*}
\end{minipage}
\begin{minipage}{.5\linewidth}
\begin{align*}
\X_i^l(t) &= \mu^l(t) + \sum_{m=1}^{\infty} \zeta_{i,m}\varPsi_m^l(t)  \nonumber  \\ 
&= \mu^l(t) + \sum_{m=1}^{M}\zeta_{i,m}\varPsi_m^l(t) +  v_i^l(t),
\end{align*}
\end{minipage}
\vspace{.1in}

\noindent where $\phi_k^e(t)$ and $\varPsi_m^l(t)$ represent the $k$\textsuperscript{th} and $m$\textsuperscript{th} functional principal components for the partially observed and remaining segments of curve on day $n+1$, respectively; $\xi_{i,k}$ and $\zeta_{i,m}$ are the principal component scores of $\X_i^e(t)$ and $\X_i^l(t)$, respectively; $K$ and $M$ denote the number of retained components, respectively; $\eta_i^e(t)$ and $v_i^l(t)$ represent the error functions associated with the partially observed data and remaining data periods, respectively, due to finite truncations.

Let $\bm{\zeta}_m = \left[\zeta_{1,m}, \cdots, \zeta_{n,m}\right]^{\top}$ and $\bm{\xi}_k = \left[\xi_{1,k}, \cdots, \xi_{n,k}\right]^{\top}$. Using $\bm{\zeta} = \left[\bm{\zeta}_1 ,\cdots, \bm{\zeta}_M\right] $ as the response variable and $\bm{\xi} = \left[\bm{\xi}_1, \cdots, \bm{\xi}_K\right]$ as a predictor, the equation considered can be expressed as
\begin{equation}
\bm{\zeta} = \bm{\xi} \times \bm{\varsigma}, \label{eq zeta}
\end{equation}
where $\bm{\varsigma}$ can be estimated by OLS, leading to 
\begin{equation}
\widehat{\bm{\varsigma}} = \left( \bm{\xi}^{\top}\bm{\xi} \right) ^{-1} \bm{\xi}^{\top} \bm{\zeta}, \label{eq varsigma}
\end{equation}
with $\bm{\xi}^{\top} \bm{\zeta}$ jointly estimated by their cross-covariance structure
\begin{equation*}
\int_{ t \in \mathcal{I}} \int_{ s \in \mathcal{I}_e} \phi_k(s) \varPsi_m(t)\cov \left[ \bm{\X}^e(s),\bm{\X}^l(t) \right]dsdt, \qquad k = 1, \cdots, K, \quad m = 1, \cdots, M, 
\end{equation*}
where $\bm{\X}^e(s) = \left[ \X_1^e(s), \cdots, \X_n^e(s) \right]^{\top} $ and $\bm{\X}^l(t) = \left[ \X_1^l(t), \cdots, \X_n^l(t) \right]^{\top} $ represent two vectors of functions that correspond to the partially observed and remaining segments of curve on day $n+1$, respectively.

From Equation~\eqref{eq flr}, we can obtain a point forecast of $\X_{n+1}^l(t)$ as
\begin{equation}
\widehat{\X}_{n+1}^l(t) = \widehat{\mu}^l(t) + \sum_{m=1}^{\infty} \zeta_{n+1,m}\varPsi_m^l(t), \label{eq flr estimate}
\end{equation}
which has an approximation based on Equation~\eqref{eq zeta} and Equation~\eqref{eq flr estimate} as
\begin{equation*}
\widehat{\X}_{n+1}^l(t) \approx \widehat{\mu}^l(t) + \bm{\widehat{\xi}_{n+1} }\times \bm{\widehat{\varsigma}} \times \widehat{\bm{\varPsi}}^l(t),
\end{equation*}
where $\widehat{\bm{\varsigma}}$ is estimated from Equation~\eqref{eq varsigma}, and $\widehat{\bm{\varPsi}}^l(t) = \left[\widehat{\varPsi}_1^l(t), \cdots, \widehat{\varPsi}_M^l(t) \right] $.

\section{Interval forecast methods}\label{sec: interval}

Prediction intervals are often used to assess the probabilistic uncertainty associated with point forecasts. As emphasised in \cite{Chatfield93}, it is essential to provide interval forecasts for the following reasons: 
\begin{inparaenum}[(1)]
\item assess future uncertainty;
\item enable different strategies to be planned for a range of possible outcomes indicated by the interval forecasts;
\item compare forecasts from different methods more thoroughly; 
\item explore different scenarios based on different assumptions.
\end{inparaenum}
It is always essential to be clear about the sources of errors before quantifying forecast uncertainty via computations. In our functional principal component regression, we identify two sources of errors:
\begin{inparaenum}[(1)]
\item the errors in estimating the regression coefficient estimates;
\item the errors remain in the model residuals.
\end{inparaenum}
In Section~\ref{subsec nonparametric pi}, we describe a nonparametric bootstrap method for constructing one-step-ahead prediction intervals for the TS method, and in Section~\ref{subsec updaing pi} we demonstrate how the prediction intervals can be updated from the BM method, PLS method, and functional linear regression.

\subsection{Nonparametric prediction interval}\label{subsec nonparametric pi}

The focus of our study is on short-term time series forecasting; we define the one-step-ahead in-sample forecast errors for estimated principal component scores as 
\begin{equation*}
\widehat{\bm{\varOmega}}_{j,k} = \widehat{\bm{\beta}}_{n-j+1,k} - \widehat{\bm{\beta}}_{n-j+1|n-j,k}, \qquad \text{for} \quad j = 1,\cdots, n-K,
\end{equation*}
where $K$ represents the number of retained principal components in Equation~\eqref{eq xt}. Sampling with replacement within these one-step-ahead forecast errors gives a bootstrap sample of $\bm{\beta}_{n+h,k}$, denoted by $\widehat{\bm{\beta}}_{n+1|n,k}^b$. For $h=1$, we have
\begin{equation*}
\widehat{\bm{\beta}}_{n+1|n,k}^b = \widehat{\bm{\beta}}_{n+1|n,k} + \widehat{\bm{\varOmega}}_{*,k}^b, \qquad \text{for} \quad b = 1, \cdots, B,
\end{equation*} 
where $\widehat{\bm{\varOmega}}_{*,k}^b$ denotes the bootstrapped forecast errors obtained by sampling with replacement from $(\widehat{\bm{\varOmega}}_{1,k}, \cdots, \widehat{\bm{\varOmega}}_{n-K,k})$, and $B=1000$ symbolises the number of bootstrap replications.

We assume the model residuals to be i.i.d random noise given the first $K$ functional principal component decomposition in Equation~\eqref{eq xt} approximates data relatively well. We can then bootstrap the model residual function $\widehat{e}_{n+1}(t)$ by randomly sampling with replacement from the historical residual functions $\{ \widehat{e}_1(t), \cdots, \widehat{e}_n(t)  \}$. Putting the two sources of errors together yields $B$ bootstrapped forecasts of $\X_{n+1}(t)$, as defined by
\begin{equation*}
\widehat{\X}_{n+1|n}^b(t) = \widehat{\mu}(t) + \sum_{k=1}^{K} \widehat{\beta}_{n+1|n,k}^b \widehat{\phi}_k(t) + \widehat{e}_{n+1}^b(t).
\end{equation*}
Therefore, $100(1-\alpha)\%$ prediction intervals can be defined as $\alpha/2$ and $(1-\alpha/2)$ empirical quantiles of $\left\lbrace \widehat{\X}_{n+1|n}^1(t), \cdots, \widehat{\X}_{n+1|n}^B(t)  \right\rbrace $. With a modification of function support range, this nonparametric prediction interval approach also works for the BM method.

\subsection{Updating prediction interval}\label{subsec updaing pi}

\subsubsection{PLS method}\label{subsec pls updating}

We can dynamically update prediction intervals using a nonparametric bootstrap method with sequentially observed new data. The initial step involves bootstrapping $B$ samples of the TS forecast regression coefficient function, denoting them by $\widehat{\bm{\beta}}_{n+1|n}^{b, \text{TS}} = \left[ \widehat{\beta}_{n+1|n,1}^{b,\text{TS}} , \cdots, \widehat{\bm{\beta}}_{n+1|n,K}^{b, \text{TS}}   \right]^{\top}$. The bootstrapped TS regression coefficient estimates will consequently lead to $\widehat{\bm{\beta}}_{n+1}^{b, \text{PLS}}$ according to Equation~\eqref{eq beta estimate}. Then we obtain $B$ replications of $\bm{\widehat{X}}_{n+1}^{b, \text{PLS}}(t_l)$ as
\begin{equation}
\bm{\widehat{X}}_{n+1}^{b, \text{PLS}}(t_l) = \bm{\widehat{\mu}}(t_l) + \sum_{k=1}^{K} \bm{\widehat{\beta}}_{n+1,k}^{b, \text{PLS}} \widehat{\bm{\phi}}_k(t_l) + \bm{\widehat{e}}_{n+1}^b(t_l). \label{eq pls updating}
\end{equation}
Therefore, $100(1-\alpha)\%$ prediction intervals for the updated forecasts can be defined as $\alpha/2$ and $(1-\alpha
/2)$ empirical quantiles of $\left\lbrace \bm{\widehat{X}}_{n+1}^{1,\text{PLS}}(t_l), \cdots, \bm{\widehat{X}}_{n+1}^{B,\text{PLS}}(t_l)\right\rbrace$.

\subsubsection{Functional linear regression}\label{subsec flr updating}

Apart from Equation~\eqref{eq pls updating}, which constructs the pointwise prediction intervals for the remaining segments of the curve on day $n + 1$, we can also use functional linear regression together with a bootstrap method to build pointwise prediction intervals. Utilising the functional linear regression in Equation~\eqref{eq flr}, we can obtain the bootstrapped forecasts $\X_{n+1}^{l,b}(t)$ as
\begin{equation}
\widehat{\X}_{n+1}^{l,b}(t) = \mu^l(t) + \int_{ s \in \mathcal{I}_e} \left[ \X_{n+1}^e(s) - \mu(s) \right] \widehat{\beta}^b (s,t) \text{ds} + \widehat{e}_{n+1}^{l,b}(t), \qquad s \in \mathcal{I}_e, \quad t \in \mathcal{I}_l, \label{eq flr updating}
\end{equation}
where $\widehat{\beta}^b(s,t)$ represents the bootstrapped regression coefficient estimates, and $\widehat{e}^{l,b}_{n+1}(t)$ stands for the bootstrapped error function associated with the remaining time period. Parameter variation in the estimation of regression coefficients is captured by $\widehat{\beta}^b (s,t)$, while $\widehat{e}_{n+1}^{l,b}(t)$ measures the model variability in the fitted model.

We assume that the one-step-ahead forecast errors do not correlate and implement the i.i.d. bootstrap method by sampling with replacement from historical errors $\left\{\widehat{e}_K^l(t), \cdots, \widehat{e}_n^l(t)\right\}$. We also use bootstrapping to obtain $\widehat{\beta}^b(s,t)$ in Equation~\eqref{eq flr updating} by sampling with replacement from the original functional time series expressed as
\begin{equation*}
\X_i^b(t) = \mu(t) + \sum_{k=1}^{\min(n,\infty)} \beta_{i,k}^b \phi_k(t), \qquad i = 1, \cdots, n,
\end{equation*}
where $\left(\beta_{1,k}^b, \cdots, \beta_{n,k}^b\right)$ denote the $k$\textsuperscript{th} bootstrapped principal component scores via maximum entropy \citep[see also][]{Shang17}. With a set of bootstrapped data $\{ \X_1^b(t), \cdots, \X_n^b(t)  \}$ available, we apply the functional linear regression of Equation~\eqref{eq flr} to obtain bootstrapped estimates of regression coefficient function.

For bootstrapping the multivariate time series $\{\bm{\beta}_1, \cdots, \bm{\beta}_n\}$ we apply the maximum entropy bootstrap method \citep{Vinod04} which has the following advantages: 
\begin{inparaenum}[(1)]
\item stationarity of principal component scores is not required;
\item time series ranks are computed by the method;
\item the bootstrap samples satisfy the ergodic theorem, central limit theorem, and mean preserving constraint;
\item the method can be applied to panel time series in which cross-covariance of the original data matrix is well preserved.
\end{inparaenum}
In practice, we utilise the maximum entropy bootstrap algorithm described in \cite{VL09}, with implementation via the \verb|meboot.pdata.frame| function of the \textit{meboot} package in R \citep{R}.

\section{Measures of point forecast accuracy}\label{sec: accuracy}

\subsection{Absolute and squared forecast errors}\label{subsec point error}

We calculate the point forecasts between the proposed methods and evaluate their forecast accuracy using mean absolute forecast error (MAFE) and mean squared forecast error (MSFE). The errors assess the absolute and squared differences between the forecasts and the actual values of the variable being forecast, with equations expressed as
\begin{align}
\text{MAFE}_j &= \frac{1}{q} \sum_{i=1}^{q} \left| \X_i(t_j) - \widehat{\X}_i(t_j) \right|, \\
\text{MSFE}_j &= \frac{1}{q} \sum_{i=1}^{q} \left[ \X_i(t_j) - \widehat{\X}_i(t_j) \right]^2,
\label{eq: loss functions}
\end{align}
where $q$ represents the number of curves in the forecasting period, $\X_i(t_j)$ represents the actual holdout sample for the $j$\textsuperscript{th} time period in the $i$\textsuperscript{th} curve, and $\widehat{\X}_i(t_j)$ represents the point forecasts for the holdout sample. To apply the updating algorithm detailed in Section~\ref{sec: updating}, at least one data point on the $i$\textsuperscript{th} curve has to be observed. Using TS, BM, PLS and BM methods the updating period can start at the $j = 2^{\text{nd}}, \cdots, 1620$\textsuperscript{th} 15-second tick interval for the CIDRs of VIX data. However, the functional linear regression method requires the updating period to start from the $j=6^{\text{th}}$ 15-second tick interval as the method is not suitable with initial zeros (added by linear interpolation, see Section~\ref{sec: VIX} for more details) on each curve.

\subsection{Mixed errors}

We calculate mean mixed error (MME) functions that are asymmetric loss functions penalising incorrect predictions. \cite{BF96}, \cite{FIK09} and \cite{KCM18} have previously considered by these measures in the evaluation of volatility forecasting methods. They are essential for option market participants as sellers (buyers) are more likely to worry about under (over)-prediction than buyers (sellers). The MME(U) is defined to penalise under-predictions more heavily while the MME(O) gives more substantial penalties to under-predictions.
\begin{align*}
\text{MME}_j(U) = \frac{1}{q} \left[ \sum_{i=i_1^O}^{i_N^O} \left| \X_i(t_j) - \widehat{\X}_i(t_j) \right| + \sum_{i=i_1^U}^{i_N^U} \sqrt{| \X_i(t_j) - \widehat{\X}_i(t_j)|}  \right] 
\end{align*}
and
\begin{align*}
\text{MME}_j(O) = \frac{1}{q} \left[ \sum_{i=i_1^O}^{i_N^O}\sqrt{| \X_i(t_j) - \widehat{\X}_i(t_j)|}  + \sum_{i=i_1^U}^{i_N^U}  \left| \X_i(t_j) - \widehat{\X}_i(t_j)\right| \right] ,
\end{align*}
where $q$ represents the number of curves in the forecasting period, $\left\{  i_1^U, \cdots, i_N^U; \ i_N^U \leq q  \right\} $ stands for the under-predictions, and $\left\{ i_1^O, \cdots, i_N^O; \ i_N^O \leq q \right\}$ stands for the over-predictions.

\subsection{Correct predictor of the direction of change}

To measure how well our models can forecast the direction of movement, we also calculate the mean correct predictor of the direction of change (MCPDC) as the percentage of $\widehat{\X}_i(t_j)$ that has the same sign as the corresponding observations $\X_i(t_j)$. MCPDC ignores the magnitude of errors and is also employed in other studies of the CBOE volatility index, for example in \cite{BG14} and \cite{KCM18}.

\subsection{Interval scores}\label{subsec interval scores}

We calculate the interval score of \cite{GR07} \citep[see also][]{GK14} to evaluate interval forecast accuracy. The one-step-ahead prediction intervals for each year in the forecasting period can be calculated at the $(1-\alpha) \times 100\%$ nominal coverage probability. We consider the common case of the symmetric $(1-\alpha) \times 100\%$ prediction interval, with lower and upper bounds serving as predictive quantiles at $\alpha/2$ and $1-\alpha/2$, denoted by $\widehat{\X}^l(t_j)$ and $\widehat{\X}^u(t_j)$. The scoring rule of \cite{GR07} for the interval forecast at time point $t_j$ is 
\begin{align*}
S_{\alpha}\left[  \widehat{\X}^l(t_j), \widehat{\X}^u(t_j); \ \X(t_j) \right] = \left[ \widehat{\X}^u(t_j)- \widehat{\X}^l(t_j) \right]  &+ \frac{2}{\alpha}\left[ \widehat{\X}^l(t_j) - \X(t_j) \right] \mathds{1} \{  \X(t_j) < \widehat{\X}^l(t_j)\} \\
&+ \frac{2}{\alpha}\left[ \X(t_j) - \widehat{\X}^u(t_j) \right] \mathds{1}\{ \X(t_j) > \widehat{\X}^u(t_j)\},
\end{align*}
where $\mathbbm{1}{\{\cdot \}}$ represents the binary indicator function, and $\alpha$ (customarily $\alpha = 0.2$) denotes the level of significance. The interval score rewards a narrow prediction interval, if and only if the true observation lies within the prediction interval. When $\X(t_j)$ is contained in the interval, and  the distance between $\widehat{\X}^l(t_j)$ and $\widehat{\X}^u(t_j)$ is minimised, the optimal interval score will be achieved. 

The mean interval score for each time point $j$ is calculated by averaging interval scores over different days in the forecasting period as
\begin{equation*}
\overline{S}_{\alpha,j} = \frac{1}{q}\sum_{i=1}^{q} S_{\alpha}\left[  \widehat{\X}_i^l(t_j), \widehat{\X}_i^u(t_j); \ \X_i(t_j) \right], \qquad j = 2, \cdots, 1620,
\end{equation*}
where $S_{\alpha} [ \widehat{\X}_i^l(t_j), \widehat{\X}_i^u(t_j); \ \X_i(t_j)] $ denotes the interval score for the $i$\textsuperscript{th} day of the forecasting period.

\section{Results}\label{sec: results}

We calculate and compare the point and interval forecast accuracies of the functional principal component regression with univariate and multivariate time series forecasting methods in Table~\ref{tab_uni_var_all}. Generally, the multivariate (VAR) time series forecasting method produces slightly larger forecast errors than the univariate time series (ARIMA) forecasting method. This result may be due to there being more parameters to be estimated in the VAR model as opposed to the ARIMA model.

\begin{table}[!htbp] 
\centering 
\tabcolsep 0.2in
\caption{Comparison of point and interval forecast accuracies between a univariate time series forecasting method and a multivariate time series forecasting method, for VIX CIDRs over the 40 days in the forecasting period.} \label{tab_uni_var_all} 
\begin{tabular}{@{\extracolsep{5pt}} lrr|rr|rr@{}} 
\toprule
& \multicolumn{2}{c}{MSFE} & \multicolumn{2}{c}{MAFE} & \multicolumn{2}{c}{Interval score} \\  \cline{2-7} 
& ARIMA & VAR & ARIMA & VAR & ARIMA & VAR \\ 
\midrule
Min & \textBF{0.7939} & 1.3486 & \textBF{0.6900} & 0.9866 & \textBF{8.2059} & 8.7041 \\ 
Median & \textBF{8.1868} & 8.6180 & \textBF{2.3820} & 2.5446 & \textBF{9.1384} & 9.2167 \\ 
Mean & \textBF{22.0173} & 25.3357 & \textBF{3.0447} & 3.3261 & \textBF{15.7863} & 17.3692 \\ 
Max & 354.7139 & \textBF{342.1089} & 16.1869 & \textBF{16.0261} & 128.4479 & \textBF{124.2109} \\ 
\bottomrule
\end{tabular} 
\end{table} 

Hereafter, the ARIMA forecasting method is used to compare point and interval forecast accuracies between the standard functional PCA, the robust functional PCA, and the RobRSVD estimation for the TS method. Table~\ref{tab ts_point_interval} demonstrates that the three methods exhibit very close point and interval forecast accuracies.

\begin{table}[!htbp] \centering 
\tabcolsep 0.02in
\caption{Comparison of point and interval forecasting accuracies of TS methods with standard functional principal component analysis, robust functional principal component analysis and robust methods utilising the regularised singular value decomposition, denoted as FPCA, M-FPCA and RobRSVD, respectively.} \label{tab ts_point_interval} 
\begin{small}
\begin{tabular}{@{\extracolsep{5pt}} lrrr|rrr|rrr@{}} 
\toprule 
& \multicolumn{3}{c}{MSFE} & \multicolumn{3}{c}{MAFE} & \multicolumn{3}{c}{Interval score} \\  \cline{2-10} 
& FPCA & M-FPCA & RobRSVD & FPCA & M-FPCA & RobRSVD & FPCA & M-FPCA & RobRSVD \\ 
\midrule
Min & 0.7939 & 0.7768 & \textBF{0.7753} & 0.6900 & 0.6774 & \textBF{0.6770} & \textBF{8.2059} & 8.2445 & 8.6548 \\ 
Median & 8.1868 & \textBF{8.0633} & 8.1196 & 2.3820 & \textBF{2.3789} & 2.3792 & 9.1384 & \textBF{9.0990} & 9.2289 \\ 
Mean & 22.0173 & \textBF{22.0134} & 22.0224 & 3.0447 & \textBF{3.0429} & 3.0465 & 15.7863 & 15.8190 & \textBF{15.6758} \\ 
Max & \textBF{354.7139} & 355.0930 & 354.9265 & 16.1869 & \textBF{16.1776} & 16.1833 & 128.4479 & 130.4321 & \textBF{127.0060} \\ 
\bottomrule 
\end{tabular} 
\end{small}
\end{table}

\subsection{Updating point forecasts}\label{subsec updating point forecasts}

\begin{figure}[!htbp]
\centering
\subfloat{{\includegraphics[width=8cm]{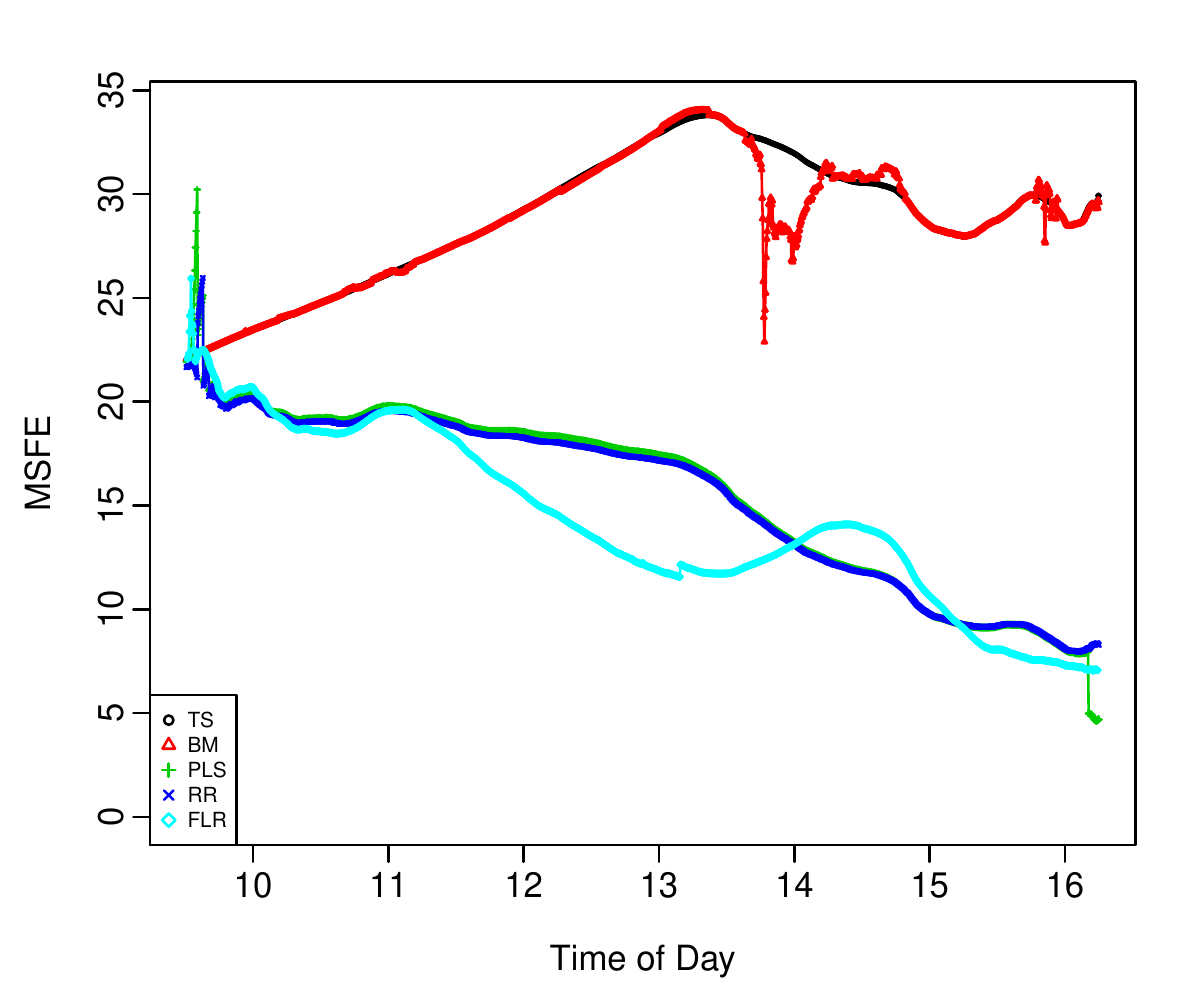} }}
\qquad
\subfloat{{\includegraphics[width=8cm]{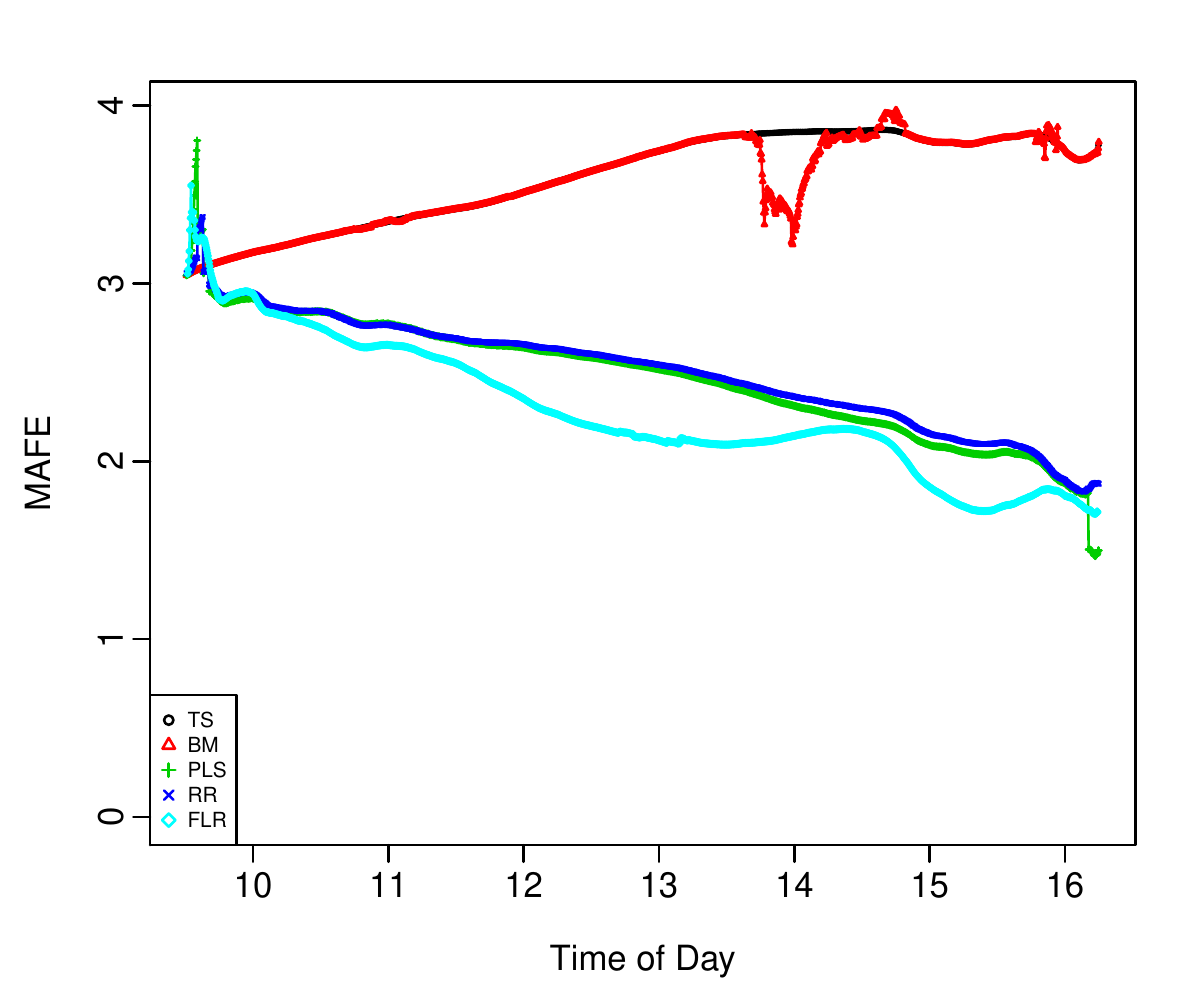} }}
\caption{A comparison of point forecast accuracy between the TS method and four dynamic updating methods, as measured by $\text{MSFE}$ and $\text{MAFE}$. Note that FLR stands for functional linear regression.}\label{fig: 3}
\end{figure}

As we receive partially observed data from the most recent curve, we can dynamically update our forecasts in the hope of achieving better forecast accuracy. Using the five dynamic updating methods detailed in Section~\ref{sec: updating}, we construct plots of averaged MSFE and MAFE across all discretised (evaluation) time points as shown in Figure~\ref{fig: 3}. When the BM method, the RR method, and functional linear regression are used, the forecast errors tend to decrease with some fluctuation. Averaged over the forecasting period and measured by the averaged MAFE, the functional linear regression has the best point forecast accuracy. Assessed by the averaged MSFE, the functional linear regression also has the lowest forecast errors across all daily trading times, except for the hour (approx.) from 13:00:00 to 15:00:00 during which time the most volatile movements of the VIX index were recorded as shown in Figure~\ref{fig: 3}. As we receive increasingly more partially observed data in the most recent curve, dynamic updating methods tend to perform much better than the TS method. These unsurprising results highlight the advantages of a functional data-analytic approach that views data as realisations of a continuous stochastic process. As the frequency of the data increases, the performance of the dynamic updating method should improve even further, as interpolation errors will likely decrease.

\cite{DB02} tests are also conducted to formally assess the most accurate forecasting method, with hypotheses and corresponding results shown in Table~\ref{tab_dm}. For example, in Test 1 we consider a squared-loss function and an absolute-loss function as defined in Equation \eqref{eq: loss functions}. The null hypothesis is that the two forecasts have the same accuracy, while the alternative hypothesis is that the FLR method is more accurate than the TS model. With the help of the \textit{forecast} package in R \citep{R}, almost zero $p$-values ($<2.2e-16$) are obtained for both loss functions, indicating that the null hypothesis should be rejected. Results in all other rows of the same table can be interpreted similarly. Overall, Table~\ref{tab_dm} shows that the functional linear regression model outperforms all other methods in producing accurate point forecasts. Such superior performance is not only sample-specific but can also draw inferences on the entire population.

\begin{table}[!htbp] \centering 
\tabcolsep 0.09in
\caption{Diebold-Mariano tests of accuracy for point forecasts obtained by the TS method and four dynamic updating methods.}\label{tab_dm} 
\begin{tabular}{@{\extracolsep{5pt}} cccccc@{}} 
\toprule
 & & \multicolumn{2}{c}{Squared-error loss} &  \multicolumn{2}{c}{Absolute-error loss} \\  \cline{3-6} 
  & Hypotheses  & Test Statistic & $p$-value & Test Statistic & $p$-value \\ 
\midrule 
 \multirow{2}{*}{Test 1} & $\text{H}_0: \text{Error}_{\text{FLR}} = \text{Error}_{\text{TS}}$  & \multirow{2}{*}{-89.0106} & \multirow{2}{*}{$<2.2e-16$} & \multirow{2}{*}{-83.9062} & \multirow{2}{*}{$<2.2e-16$} \\ 
 & $\text{H}_A: \text{Error}_{\text{FLR}} < \text{Error}_{\text{TS}}$ &  &  &  & \\
 \hline
 \multirow{2}{*}{Test 2} & $\text{H}_0: \text{Error}_{\text{FLR}} = \text{Error}_{\text{BM}}$  & \multirow{2}{*}{-88.7233} & \multirow{2}{*}{$<2.2e-16$} & \multirow{2}{*}{-83.3199} & \multirow{2}{*}{$<2.2e-16$} \\
 & $\text{H}_A: \text{Error}_{\text{FLR}} < \text{Error}_{\text{BM}}$ &  &  &  & \\
 \hline
 \multirow{2}{*}{Test 3} & $\text{H}_0: \text{Error}_{\text{FLR}} = \text{Error}_{\text{PLS}}$  & \multirow{2}{*}{-24.1949} & \multirow{2}{*}{$<2.2e-16$} & \multirow{2}{*}{-43.4326} & \multirow{2}{*}{$<2.2e-16$} \\
 & $\text{H}_A: \text{Error}_{\text{FLR}} < \text{Error}_{\text{PLS}}$ &  &  &  & \\
 \hline
 \multirow{2}{*}{Test 4} & $\text{H}_0: \text{Error}_{\text{FLR}} = \text{Error}_{\text{RR}}$  & \multirow{2}{*}{-21.7857} & \multirow{2}{*}{$<2.2e-16$} & \multirow{2}{*}{-49.3449} & \multirow{2}{*}{$<2.2e-16$} \\
 & $\text{H}_A: \text{Error}_{\text{FLR}} < \text{Error}_{\text{RR}}$ &  &  &  & \\
\bottomrule
\end{tabular} 
\end{table} 

Forecast results reported so far are obtained for the VIX index data from January to June of 2017. To add robustness to the results, we have also conducted an extended analysis using all of 2016 (details included in the Supplement), providing additional support for the proposition that functional time series forecasting models with dynamic updating (especially the FLR model) should be considered in order to produce accurate forecasts of the VIX index. Furthermore, it also indicates that our proposed models do not depend on a particular data range to provide the desired results outlined here.

\subsection{Updating interval forecasts}\label{subsec updating interval forecasts}

\begin{figure}[!htbp]
\centering
\includegraphics[width=12cm]{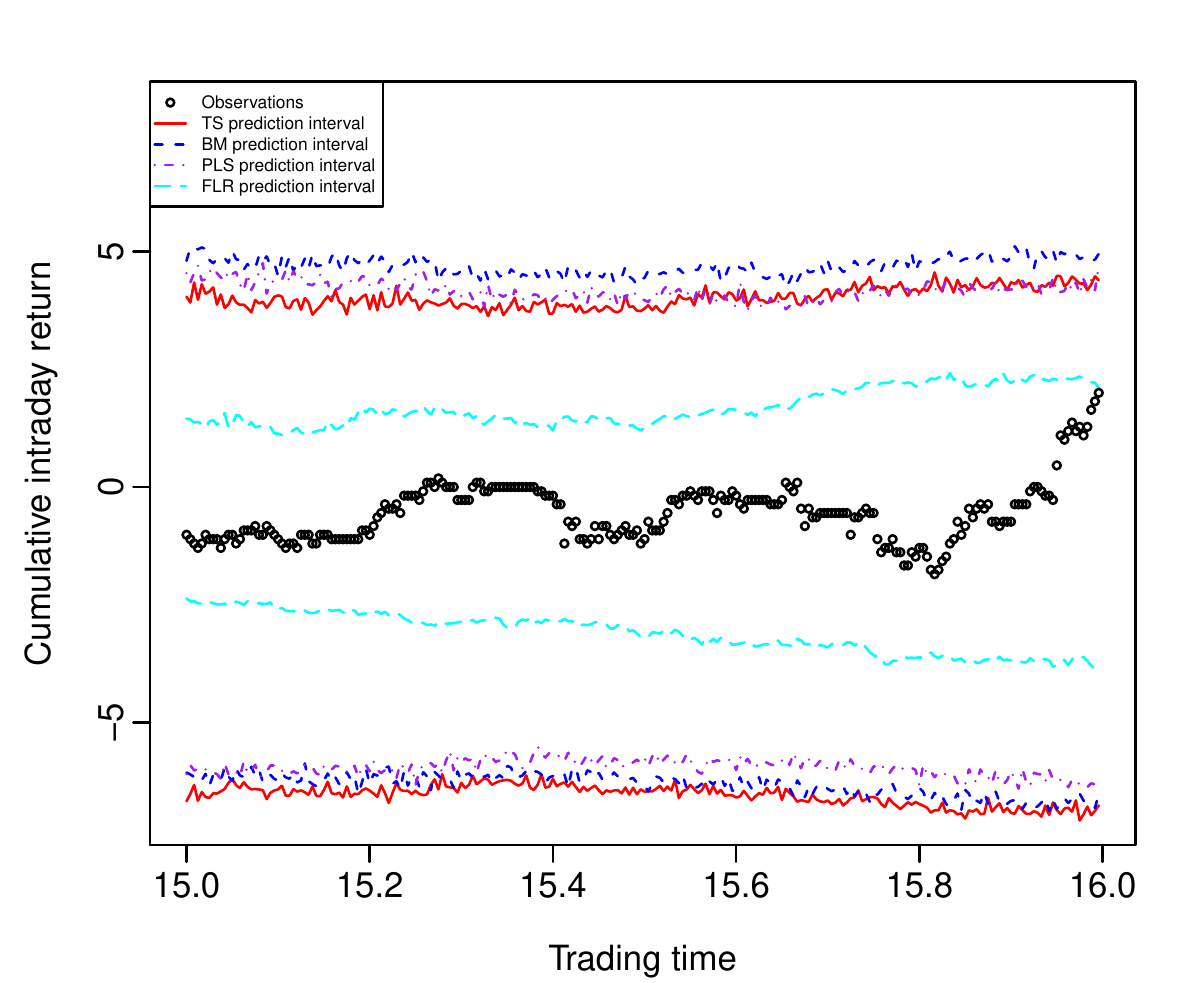}
\caption{Updating 80\% pointwise prediction intervals using the TS method, BM method, PLS method, and functional linear regression, respectively.}\label{fig 4}
\end{figure}

After observing the VIX intraday cumulative returns from 9:30:00 to 14:59:45 on 30 June 2017, the BM, PLS, and the functional linear regression method can be applied to dynamically update the interval forecasts for the remaining trading time period of the same day. Figure \ref{fig 4} shows the 80\% pointwise prediction intervals for the next hour using three dynamic updating methods, compared with the TS model that does not incorporate updating. It can be seen that the functional linear regression produces the narrowest prediction interval, leading to the most accurate evaluation of forecast uncertainty.

\subsection{Economic evaluation of forecasts} \label{subsec trading strategy}

Building on the superior predictability of dynamic functional time series models, we consider applications that highlight the economic value of adopting the proposed forecasting methods. As the VIX index itself cannot be directly traded, implementing a trading strategy involves the adoption of intraday data for a related derivative product. To the best of our knowledge, there are no existing intraday VIX trading strategies outlined in the literature, \footnote{For instance, \cite{KST08}, \cite{KS11} and \cite{Kourtis16} only consider trading strategies that were utilising either daily or monthly futures data.}.

\begin{figure}[!htbp]
  \centering
  \subfloat{{\includegraphics[width=6cm]{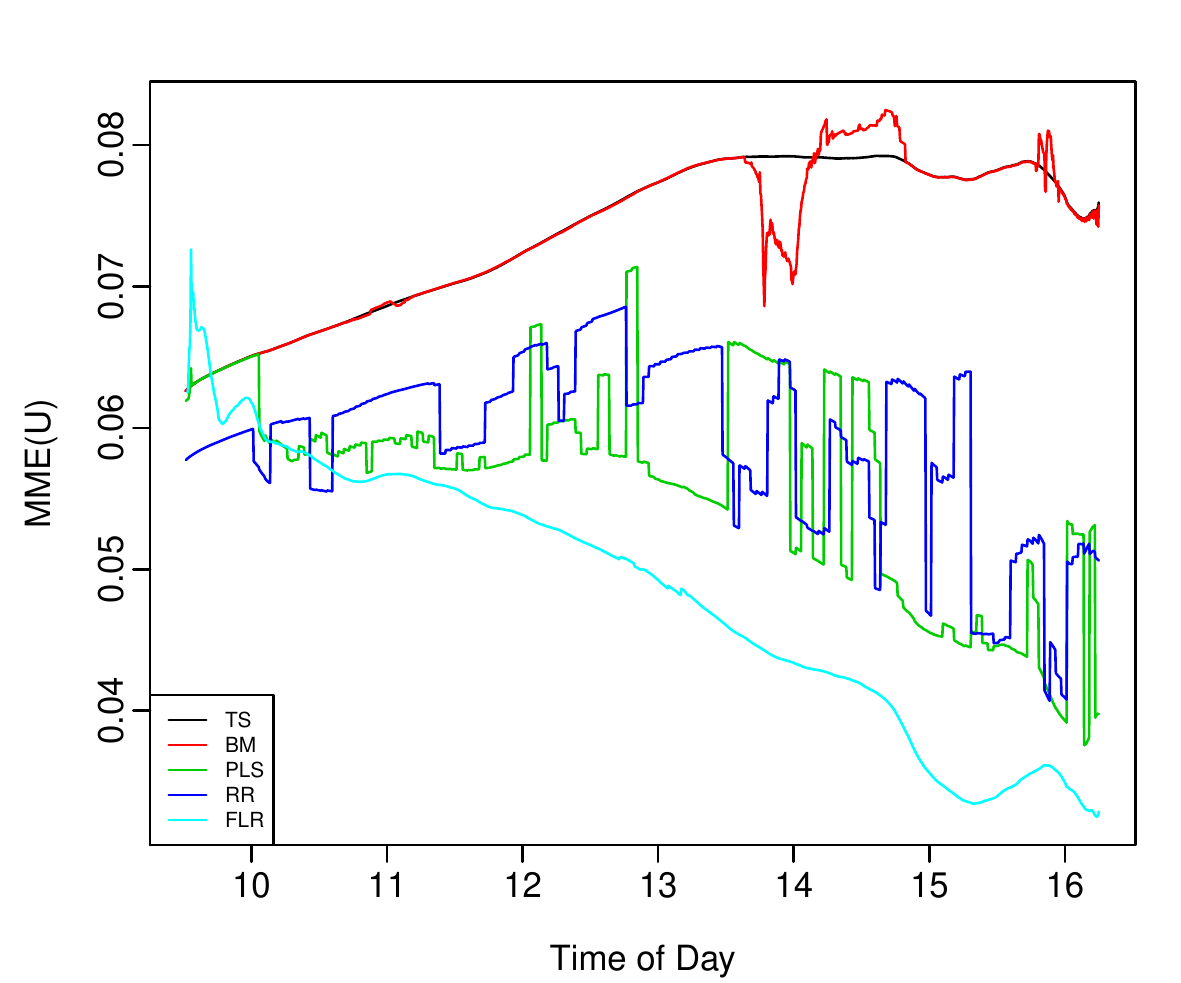} }}
  \subfloat{{\includegraphics[width=6cm]{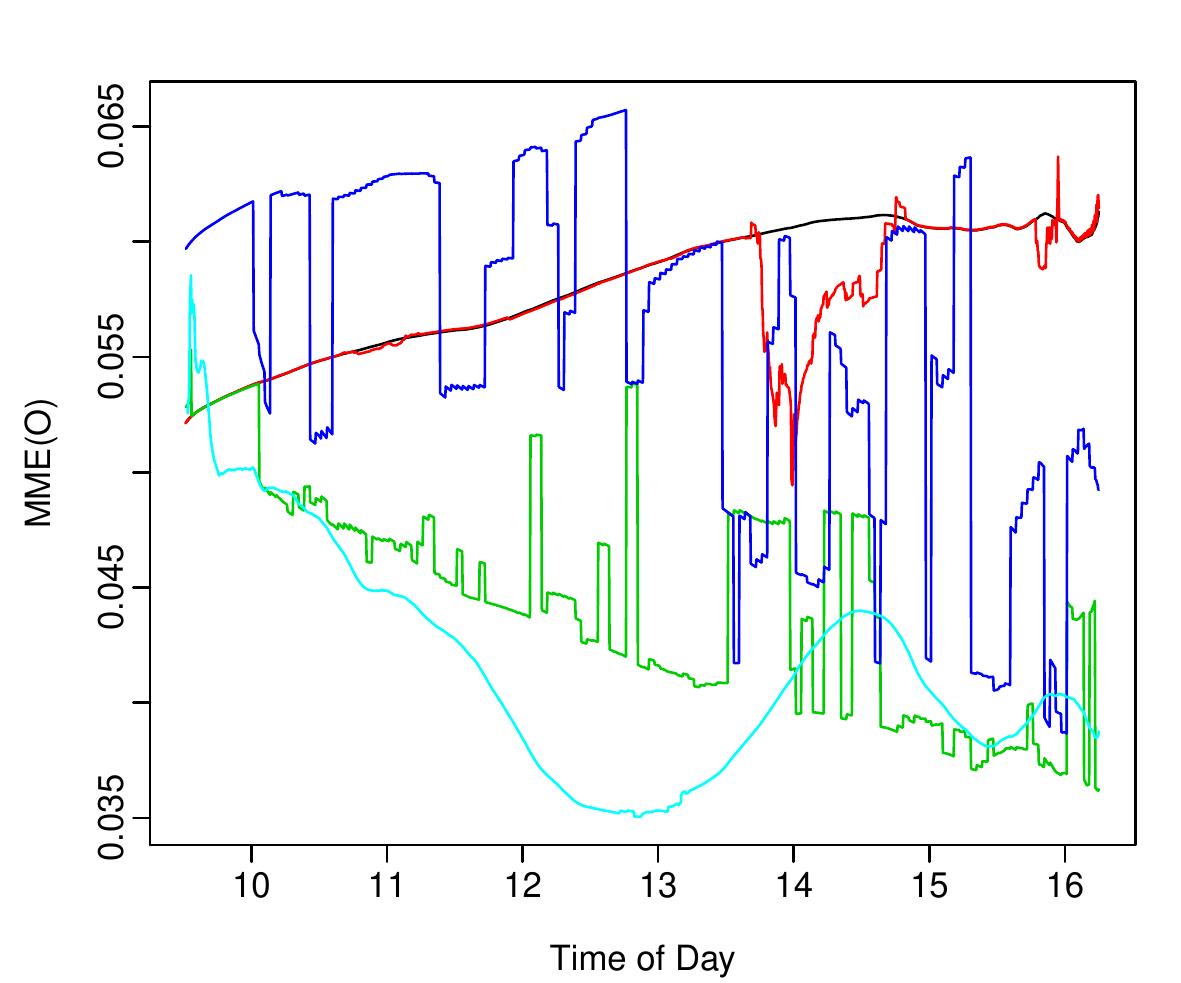} }} 
  \subfloat{{\includegraphics[width=6cm]{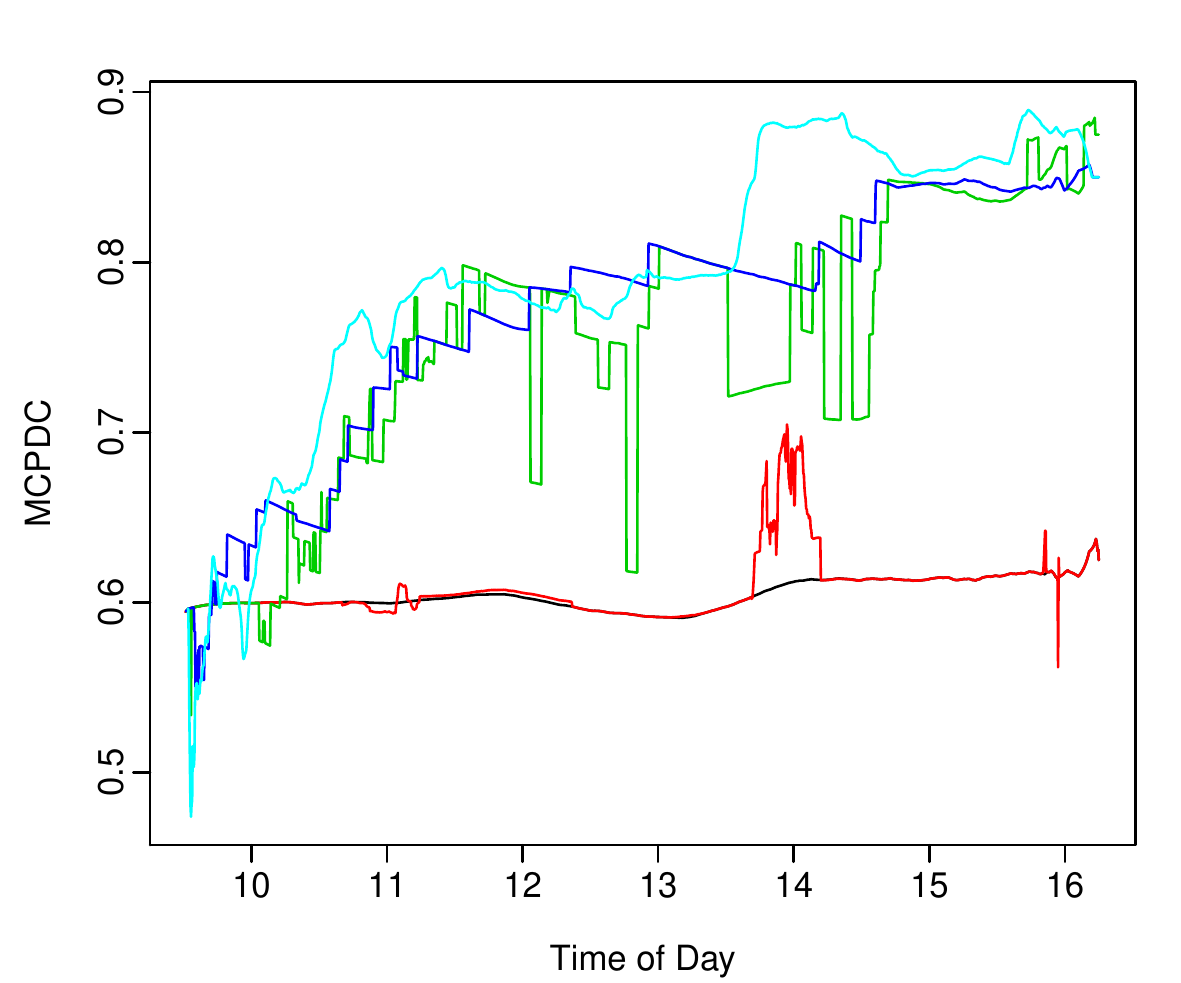} }}
  \caption{Mean Mixed Error (Under), labelled as MME(U); Mean Mixed Error (Over), labelled as MME(O); and Mean Correct Predictor of the Direction of Change, labelled as MCPDC, computed for the TS method and four dynamic updating methods.}\label{fig 5}
\end{figure}

Utilising the MME lose functions and MCPDC measure, an indication of how well our VIX forecasts at providing trading signals is presented in Figure~\ref{fig 5}. Our proposed functional time series methods all have MCPDC\footnote{MCPDC of TS: max 63.75\%, mean 60.48\%; MCPDC of BM: max 70.47\%, mean 60.88\%; MCPDC of PLS: max 88.50\%, mean 75.22\%; MCPDC of RR: max 85.75\% mean 76.98\%; MCPDC of FLR: max 88.95\% mean 79.13\%.} values that are better than the commonly used benchmark. By benchmark, we mean that a random model should correctly predict the direction of change of the index 50\% of the time. Among our functional time series methods, the FLR model produces the highest mean MCPDC of about 80\%. This result indicates that our considered functional time series approaches should perform well when adapted to provide a buy/sell indicator to pursue long/short positions in VIX index derivative products. Furthermore, the Mean Mixed Error (Under/Over) results indicate that in general there is no significant bias arising from our approaches, i.e., the forecasts do not systematically under or over predict the VIX index level. Such a concern would be particularly relevant for practitioners with stop-loss limits, or those informing the purchase of VIX-related options using our VIX index forecasts.

However, many practical considerations need to be accounted for when implementing any trading strategy. These factors are particularly numerous in a high-frequency environment, such as ours. Many are idiosyncratic to individual market participants including, market frictions, slippage, transaction costs, order book depth, microstructure noise, order latency, exchange rebates, stop-loss limits, and leverage and portfolio diversification concerns. Without all knowledge of such factors, outlining a simulated trading strategy would prove to be unrealistic.

\section{Conclusion}\label{sec: conclusion}

Functional time series forecasting and updating methods discussed here treat observed data as realisations of a continuous stochastic process. We use functional PCA to model the temporal dependence among functional curves, reducing the dimensionality of data. A set of functional principal components that explains at least 90\% of the total variation in the 15-second VIX data is considered so that the main features of the original functional time series can be maintained. We produce one-day-ahead forecast VIX CIDRs by forecasting $K$ retained principal component scores through a univariate or multivariate time series forecasting technique. In particular, the forecasts are obtained by multiplying the forecast principle component scores by the estimated functional principal components and then adding the estimated mean function. 

We consider five dynamic updating methods, which utilise partially observed data for the most recent curve to improve forecast accuracy. The BM method applies the general structure of the TS method to our VIX data but rearranges the function support to update forecasts during the trading day. The OLS method treats the partially observed data in the most recent curve as a discrete response vector and regresses it against the corresponding discretised functional principal components. However, the singularity problem may plague the implementation of the OLS method, when the number of partially observed data points is less than the number of functional principal components. To overcome this problem, we consider methods that shrink the estimated regression coefficients. The RR method shrinks the regression coefficient estimates towards zero, whereas the PLS method shrinks the regression coefficient estimates towards $\bm{\hat{\beta}}_{n+1|n}^{\text{TS}}$. The amount of shrinkage in the RR and PLS methods is controlled by a shrinkage parameter $\lambda$ that can be tuned by minimising the forecast error measure within a holdout data sample. One commonality between the OLS, RR, and PLS methods is that they attempt to update forecasts by utilising discrete time points. Conversely, the functional linear regression first decomposes two blocks of functional time series that correspond to the partially observed data and remaining data periods via the Karhunen-Lo\`{e}ve expansion. The method then forms a linear model based on the two sets of principal component scores via the ordinary least squares method and estimates the regression coefficient function from which the updated forecasts can be obtained. Functional linear regression outperforms all other dynamic updating methods investigated in producing the most accurate point forecasts assessed using averaged MSFE and MAFE over the different discretised time points in the testing sample. 

As a means of assessing forecast uncertainty, a non-parametric bootstrap method is used to construct prediction intervals for the TS and BM methods. The PLS prediction intervals are updated as we sequentially observe new data in the most recent curve. When updating prediction intervals through functional linear regression, we considered the maximum entropy bootstrapping and the non-parametric bootstrapping method to obtain bootstrapped regression coefficient functions and bootstrapped error functions, respectively. These approaches result in improved interval forecast accuracy, allowing us to compare the different functional time series methods considered.

The focus of this paper is on statistical forecasting accuracy. However, the real-time dynamically updating VIX forecasting results bring with them a number of important economic implications. Firstly, as shown by \cite{PS16}, implied volatility index forecasts can feed into profitable trading strategies involving tradable VIX derivatives, which despite many additional considerations, could also be adopted by practitioners in our high frequency environment. Secondly, considering prediction intervals as we do, boasts the advantage of enabling a number of different strategies to be planned and specified for a range of possible scenarios \citep{Chatfield93}. Lastly, demonstrating predictability in the `fear gauge' is of great benefit to market analysts and practitioners alike, as it enables them to better prepare for the impact of future market movements. For these reasons, economic evaluations of hedging and trading strategies based on the proposed functional time series methods could form the basis for future research.

\newpage
\bibliographystyle{agsm}
\bibliography{VIX}

\end{document}